\documentclass[12pt,doublespacing]{article}
\usepackage{fullpage}
\usepackage{setspace}
\usepackage{authblk}
\usepackage{cite}
\usepackage[title]{appendix}
\usepackage{amssymb}
\usepackage{graphicx}
\usepackage{graphics}
\usepackage[cmex10]{amsmath}
\usepackage{latexsym,amsfonts}
\usepackage{amsthm}
\usepackage{subfigure}
\usepackage{hyperref} 
\usepackage{enumitem}

\begin{document}
\title{A Time Domain Volume Integral Equation Solver to Analyze Electromagnetic Scattering from Nonlinear Dielectric Objects}
\author[1]{Sadeed Bin Sayed}
\author[2]{Rui Chen} 
\author[3]{Huseyin Arda Ulku} 
\author[4]{Hakan Bagci} 

\affil[1]{Sadeed Bin Sayed was with the Electrical and Computer Engineering Program, Computer, Electrical, and Mathematical Science and Engineering Division, King Abdullah University of Science and Technology (KAUST), Thuwal 23955-6900, Saudi Arabia. He is now with Halliburton Far East Pointe Ltd., 639940, Singapore (e-mail: sadeed.sayed@kaust.edu.sa).\vspace{0.25cm}}
\affil[2]{Rui Chen was with the Electrical and Computer Engineering
Program, Computer, Electrical, and Mathematical Science and Engineering
Division, King Abdullah University of Science and Technology (KAUST),
Thuwal 23955-6900, Saudi Arabia. He is now with the Department of Electronics and Telecommunications, Politecnico di Torino, 10129 Turin, Italy (e-mail: rui.chen@kaust.edu.sa).\vspace{0.25cm}}
\affil[3]{Huseyin Arda Ulku is with the Department of Electronics Engineering, Gebze Technical University, Kocaeli 41400, Turkey (e-mail: haulku@gtu.edu.tr).\vspace{0.25cm}}
\affil[4]{Hakan Bagci is with the Electrical and Computer Engineering Program, Computer, Electrical, and Mathematical Science and Engineering Division, King Abdullah University of Science and Technology (KAUST), Thuwal 23955-6900, Saudi Arabia (e-mail: hakan.bagci@kaust.edu.sa).}
\footnotetext[1]{This work was supported by the King Abdullah University of Science and Technology (KAUST) Office of Sponsored Research (OSR) under Award No 2019-CRG8-4056.}
\date{}
\maketitle

\newpage
\begin{abstract}
A time domain electric field volume integral equation (TD-EFVIE) solver is proposed for analyzing electromagnetic scattering from dielectric objects with Kerr nonlinearity. The nonlinear constitutive relation that relates electric flux and electric field induced in the scatterer is used as an auxiliary equation that complements TD-EFVIE.  The ordinary differential equation system that arises from TD-EFVIE’s Schaubert-Wilton-Glisson (SWG)-based discretization is integrated in time using a predictor-corrector method for the unknown expansion coefficients of the electric field. Matrix systems that arise from the SWG-based discretization of the nonlinear constitutive relation and its inverse obtained using the Padé approximant are used to carry out explicit updates of the electric field and the electric flux expansion coefficients at the predictor and the corrector stages of the time integration method. The resulting explicit marching-on-in-time (MOT) scheme does not call for any Newton-like nonlinear solver and only requires solution of sparse and well-conditioned Gram matrix systems at every step. Numerical results show that the proposed explicit MOT-based TD-EFVIE solver is more accurate than the finite-difference time-domain method that is traditionally used for analyzing transient electromagnetic scattering from nonlinear objects.\par
\end{abstract}

\newpage

\section{Introduction}
\label{sec:introduction}
Electromagnetic and optical devices often rely on materials that exhibit strong Kerr nonlinearity~\cite{jin2011theory, boyd2020nonlinear, saleh2007fundamentals} to induce various interesting physical phenomena, such as higher-order harmonic generation~\cite{weerawarne2015}, self-focusing~\cite{zhang1998}, self-phase modulation~\cite{karlsson1991}, four-wave mixing~\cite{jin2016}, and three-state switching~\cite{amin2014}, in their response to electromagnetic excitation. The dielectric permittivity of these nonlinear materials is mathematically modeled as a power series of the magnitude of the electric field weighted with susceptibility coefficients~\cite{jin2011theory, boyd2020nonlinear, saleh2007fundamentals}.

Due to this complicated dependence of the dielectric permittivity on the electric field, design of the nonlinear electromagnetic and optical devices, even when they have simplistic geometries, has to be carried out using an electromagnetic simulator. These simulators often rely on time domain techniques since frequency domain methods operate under the assumption of time-harmonic excitation and can not be used in the presence of strong nonlinearities~\cite{jin2011theory}.

Majority of the time domain solvers developed to simulate electromagnetic field interactions on materials with Kerr nonlinearity are based on the finite-difference time-domain (FDTD) method~\cite{taflove2005computational, teimoori2018, maksymov2011comparative, sarris2011, pinto2006fdtd, maksymov2006fdtd, fujii2005, van1999, joseph1997, ziolkowski1997incorporation, tran1995photonic}. This can be explained by the fact that FDTD methods are relatively straightforward to be implemented and they permit easy incorporation of the nonlinear permittivity function through the use of an auxiliary equation~\cite{taflove2005computational, maksymov2011comparative, teixeira2008}. This approach calls for iterative updates between the Maxwell equations and this auxiliary equation, which is often implemented using a Newton-type nonlinear solver or by simple explicit updates of the variables in the equations~\cite{taflove2005computational, teixeira2008}. 

Time marching schemes that rely on similar approaches have also been adopted into time domain finite element method (TD-FEM)~\cite{anees2020, abraham2019b, abraham2020, abraham2019a, yan2016, fisher2007}. Unlike FDTD, TD-FEM is not restricted to uniform grids for spatial discretization, and therefore can be used in the simulation of devices with complicated geometries without loss of accuracy and/or efficiency. That said, both FDTD and TD-FEM suffer from several well-known drawbacks of the differential equation solvers: They require the computation domain to be truncated using absorbing boundary conditions or perfectly matched layers, their accuracy is limited by numerical phase dispersion, and their time step size is often restricted by the Courant-Friedrichs-Lewy (CFL) condition~\cite{jin2011theory, taflove2005computational}.

Time domain volume integral equation (TD-VIE) solvers~\cite{sarkar2000, rao2002, jarro2012, gres2001, shanker2004, kobidze2005, liu2016, yilmaz2005, bagci2007, sayed2015, hu2016, sayed2014stable, sayed2020} do not suffer from these drawbacks of FDTD and TD-FEM. This is because they rely on a formulation where the scattered electromagnetic field is represented as a spatio-temporal convolution between the Green function of the background medium and current/field induced in the geometry. However, this convolution operation (in discretized form) is also the reason why computational cost and memory requirements of classical marching-on-in-time (MOT)-based TD-VIE solvers are high~\cite{shanker2004, kobidze2005, liu2016, yilmaz2005, bagci2007}. Furthermore, inaccurate discretization/computation of the retarded-time integrals relevant to this convolution cause late-time instabilities in the solution~\cite{sayed2015}. The former issue has been addressed by the development of plane wave time domain (PWTD) algorithm~\cite{shanker2004, kobidze2005, liu2016} and fast Fourier transform (FFT)-based schemes~\cite{yilmaz2005, bagci2007}. Late-time instability issue has been mostly alleviated by using highly-accurate interpolation functions in the temporal discretization~\cite{weile2004, wildman2004, knab1979, sayed2015, glaser2009new}.

These developments have significantly increased the range of TD-VIE solvers' applicability and enabled their use in transient analysis of electromagnetic scattering from electrically large~\cite{liu2016, yilmaz2005, bagci2007}, lossy~\cite{shanker2004}, dispersive~\cite{kobidze2005}, and/or high-contrast~\cite{sayed2015} scatterers. On the other hand, development of TD-VIE solvers for scatterers with nonlinear permittivity has been limited to only two-dimensional (2D) problems~\cite{wang2000time}. This can be explained by the fact that this 2D method uses an implicit MOT scheme and therefore requires a Newton-like nonlinear solver at every time step, which limits its computational efficiency and applicability to three-dimensional (3D) problems.

In this work, an explicit MOT-based time domain electric field volume integral equation (TD-EFVIE) solver is proposed for transient analysis of electromagnetic scattering from 3D nonlinear dielectric objects with Kerr nonlinearity. The proposed method represents the scattered electric field in the form of a (volumetric) convolution between the time domain Green function of the background medium and two unknowns, namely the total electric field and the total electric flux induced in the scatterer. Then, the time derivative of the fundamental field relation, i.e., the sum of the incident and the scattered fields is equal to the total field, is enforced in the scatterer. This yields TD-EFVIE. The nonlinear constitutive relationship between the two unknowns is used as an auxiliary equation that complements TD-EFVIE. 

To numerically solve TD-EFVIE, the scatterer is discretized into a mesh of tetrahedrons. The electric flux and the electric field are spatially expanded using ``full'' and ``half'' Schaubert-Wilton-Glisson (SWG) basis functions~\cite{schaubert1984tetrahedral, shanker2004, kobidze2005, hu2016} defined on these tetrahedrons. Inserting these expansions into TD-EFVIE and testing the resulting equation using half SWG functions yield a system of ordinary differential equations (ODEs) in time-dependent expansion coefficients of the SWG basis functions. A predictor-corrector scheme, more specifically a $PE(CE)^m$ scheme, is used to integrate this system of ODEs in time for the unknown electric field expansion coefficients~\cite{sayed2020, sayed2014stable, ulku2013, chen2019, chen2021, glaser2009new, hairer2010}. Here, $PE$ and $CE$ refer to the predictor and corrector stages, and $m$ is the number of corrector updates. Similarly, expansions of the electric field and the electric flux are inserted into the nonlinear constitutive relation and its ``inverse'' obtained using the Padé approximant~\cite{oskooi2010Meep, baker1996, pozzi1994}. The resulting equations are tested using full SWG functions at discrete time steps. This yields matrix systems that relate the electric flux expansion coefficients to those of the electric field. These matrix equations are used at PE and CE stages of the $PE(CE)^m$ scheme during the time integration for the explicit updates of the expansion coefficients. This approach does not call for Newton-like nonlinear solvers as done in the implicit MOT solvers. Even though it requires linear matrix system solutions at every time step, the Gram matrices associated with these systems are always sparse and well-conditioned (independent of the time step size). Therefore, these systems are efficiently solved using an iterative solver. 

The accuracy and the applicability of the resulting explicit MOT-based TD-EFVIE solver are demonstrated using several numerical examples. These results clearly show that the proposed method is more accurate than FDTD that is traditionally used for analyzing electromagnetic scattering from nonlinear objects. Note that preliminary versions of the method proposed in this work have been described in~\cite{ulku2015explicit, sayed2017explicit} as conference contributions.

The remainder of this paper is organized as follows. Sections~\ref{sec:tdefvie} and~\ref{sec:cons_relation} present the formulation of TD-EFVIE and the nonlinear constitutive relation and their discretization. Section~\ref{sec:pece} describes the $PE(CE)^m$ scheme that is used for the solution of the discretized TD-EFVIE (in the form of ODE systems) and the discretized nonlinear constitutive relation. This is followed by Section~\ref{sec:comments} where several comments about the proposed explicit MOT-based TD-EFVIE solver are provided. Section~\ref{sec:numres} presents numerical results that demonstrate the accuracy, the stability, and the applicability of the proposed method. Finally, Section~\ref{sec:conc} summarizes this work and outlines future research directions.

\section{Formulation}\label{sec:formulation}
\subsection{TD-EFVIE}\label{sec:tdefvie}

Let $V$ denote the volumetric support of a scatterer that resides in an unbounded background medium with permittivity $\varepsilon_{0}$ and permeability $\mu_{0}$. Electric field $\mathbf{E}^{\mathrm{inc}}(\mathbf{r}, t)$ is incident on the scatterer. It is assumed that $\mathbf{E}^{\mathrm{inc}}(\mathbf{r}, t)$ is vanishingly small for $\forall \mathbf{r} \in V$ and $t \leq 0$, and is essentially band-limited to frequency $f_{\mathrm{max}}$. In response to this excitation, the equivalent volumetric electric current $\mathbf{J}(\mathbf{r}, t)$ is induced in $V$, and $\mathbf{J}(\mathbf{r}, t)$ generates the scattered electric field $\mathbf{E}^{\mathrm{sca}}(\mathbf{r}, t)$. One can express $\mathbf{E}^{\mathrm{sca}}(\mathbf{r}, t)$ in terms of $\mathbf{J}(\mathbf{r}, t)$ using~\cite{jin2011theory}
\begin{equation}
\label{eq:sca_field}
\mathbf{E}^{\mathrm{sca}}(\mathbf{r}, t)=-\frac{\mu_{0}}{4 \pi} \int_{V} \frac{\partial_{t} \mathbf{J}(\mathbf{r}^{\prime},t^{\prime})\big|_{t^{\prime}=t-R / c_{0}}}{R} d v^{\prime}
+\frac{1}{4 \pi \varepsilon_{0}} \nabla \int_{V} \int_{-\infty}^{t-R / c_{0}} \frac{\nabla^{\prime} \cdot \mathbf{J}(\mathbf{r}^{\prime}, t^{\prime})}{R} d t^{\prime} d v^{\prime}
\end{equation}
where $\partial_{t}$ denotes the derivative with respect to time, $R=\left|\mathbf{r}-\mathbf{r}^{\prime}\right|$ is the distance between the source point $\mathbf{r}^{\prime}$ and the observation point $\mathbf{r}$, and $c_{0}=1 / \sqrt{\varepsilon_{0} \mu_{0}}$ is the speed of light in the background medium. In $V$, $\mathbf{J}(\mathbf{r}, t)$ is expressed in terms of the total electric field $\mathbf{E}(\mathbf{r}, t)$ and the electric flux $\mathbf{D}(\mathbf{r}, t)$ using
\begin{equation}
\label{eq:polarization}
\mathbf{J}(\mathbf{r}, t)=\partial_{t} \mathbf{D}(\mathbf{r}, t)-\varepsilon_{0} \partial_{t} \mathbf{E}(\mathbf{r}, t),\,\mathbf{r} \in V.
\end{equation}
Here, $\mathbf{E}(\mathbf{r}, t)$ satisfies the fundamental field relation: 
\begin{equation}
\label{eq:fund_field}
\mathbf{E}(\mathbf{r}, t)=\mathbf{E}^{\mathrm{inc}}(\mathbf{r}, t)+\mathbf{E}^{\mathrm{sca}}(\mathbf{r}, t).
\end{equation}
Inserting~\eqref{eq:sca_field} and~\eqref{eq:polarization} into the time derivative of~\eqref{eq:fund_field} for $\mathbf{r} \in V$ yields the time derivative form of TD-EFVIE [in unknowns $\mathbf{E}(\mathbf{r}, t)$ and $\mathbf{D}(\mathbf{r}, t)$]~\cite{gres2001,shanker2004,kobidze2005, yilmaz2005, bagci2007,jarro2012,sayed2015,hu2016} as
\begin{equation}
\label{eq:td_efvie}
\partial_{t} \mathbf{E}(\mathbf{r}, t)+\mathcal{L}[\mathbf{E}](\mathbf{r}, t)-\frac{1}{\varepsilon_{0}}\mathcal{L}[\mathbf{D}](\mathbf{r}, t)=\partial_{t} \mathbf{E}^{\mathrm{inc}}(\mathbf{r}, t), \quad \mathbf{r} \in V.
\end{equation}
Here, $\mathcal{L}[\mathbf{X}](\mathbf{r},t)$ is the volume integral operator defined as 
\begin{equation}
\label{eq:vie_operator}
\mathcal{L}[\mathbf{X}](\mathbf{r}, t)=-\frac{\varepsilon_{0} \mu_{0}}{4 \pi} \int_{V}\frac{1}{R}{\big.\partial_{t^{\prime}}^{3} \mathbf{X}(\mathbf{r}^{\prime}, t^{\prime})\big|_{t^{\prime}=t-R/c_{0}}} d v^{\prime}
+\frac{1}{4 \pi} \nabla \int_{V}\frac{1}{{R}}{\nabla^{\prime} \cdot \big.\partial_{t^{\prime}} \mathbf{X}(\mathbf{r}^{\prime}, t^{\prime})\big|_{t^{\prime}=t-R/c_{0}}}d v^{\prime}.
\end{equation}
Note that an additional time derivative is applied to~\eqref{eq:fund_field} to obtain~\eqref{eq:td_efvie} because this equation is in the form of an ODE in time and a $PE(CE)^m$ scheme be used to integrate it in time for the unknown $\mathbf{E}(\mathbf{r},t)$~\cite{sayed2020, sayed2014stable, ulku2013, chen2019, chen2021, glaser2009new, hairer2010}.

To discretize~\eqref{eq:td_efvie}, first $V$ is divided into a mesh of tetrahedrons, and $\mathbf{D}(\mathbf{r}, t)$ and $\mathbf{E}(\mathbf{r}, t)$ are discretized using SWG functions that are defined on triangular patches of this mesh~\cite{schaubert1984tetrahedral}. $\mathbf{D}(\mathbf{r}, t)$ is approximated using 
\begin{equation}
\label{eq:d_expansion}
\mathbf{D}(\mathbf{r}, t)=\sum_{n=1}^{N^{\mathrm{D}}}\{\mathbf{I}^{\mathrm{D}}(t)\}_{n} \mathbf{f}_{n}^{\mathrm{D}}(\mathbf{r})
\end{equation}
where $\mathbf{f}_{n}^{\mathrm{D}}(\mathbf{r})$ represents the SWG basis function set, $\{\mathbf{I}^{\mathrm{D}}(t)\}_{n}=I_{n}^{\mathrm{D}}(t)$ are the unknown time-dependent expansion coefficients, and $N^{\mathrm{D}}$ is the total number of patches in the tetrahedral mesh. $\mathbf{f}_{n}^{\mathrm{D}}(\mathbf{r})$ associated with triangular patch $S_n$ is defined as~\cite{schaubert1984tetrahedral}
\begin{equation}
\label{eq:swg_def}
\mathbf{f}_n^{\mathrm{D}}(\mathbf{r})=\left\{\begin{aligned} &\mathbf{f}^+_n(\mathbf{r})=\frac{|S_n|}{3|V^+_n|}(\mathbf{r}-\mathbf{r}^+_n), \quad \mathbf{r}\in V_n^+\\
&\mathbf{f}^-_n(\mathbf{r})=-\frac{|S_n|}{3|V^-_n|}(\mathbf{r}-\mathbf{r}^-_n), \quad \mathbf{r}\in V_n^-\\
&0, \; \mathrm{otherwise}.  \end{aligned} \right.
\end{equation}
Here, $V^{\pm}_n$ are the tetrahedrons on the two sides of $S_n$, $|S_n|$ represents the area of $S_n$, $|V^{\pm}_n|$ are the volumes of $V^{\pm}_n$, and $\mathbf{r}_n^{\pm}$ are the ``free’’ nodes of $V^{\pm}_n$, i.e., $\mathbf{r}^+_n \in V^+_n$, $\mathbf{r}^+_n \notin S_n$ and $\mathbf{r}^-_n \in V^-_n$, $\mathbf{r}^-_n \notin S_n$. Note that if $S_n$ is located on the surface of $V$, there is only one tetrahedron attached to it and $\mathbf{f}_{n}^{\mathrm{D}}(\mathbf{r})$ is set to the ``half'' SWG function $\mathbf{f}_n^{+}(\mathbf{r})$ defined in this single tetrahedron represented by $V_{n}^{+}$ in the first row of~\eqref{eq:swg_def}. Note that this combination of full and half SWG functions used in basis set $\mathbf{f}_{n}^{\mathrm{D}}(\mathbf{r})$ ensures that the normal component of $\mathbf{D}(\mathbf{r}, t)$ across any two tetrahedrons in $V$ is continuous and the normal component of $\mathbf{D}(\mathbf{r}, t)$ on the surface of $V$ is properly accounted for.

On the other hand, $\mathbf{E}(\mathbf{r}, t)$ should be approximated using a basis set that allows its normal component to be discontinuous across any two tetrahedrons in $V$. Therefore, $\mathbf{E}(\mathbf{r}, t)$ is expanded using half SWG basis functions~\cite{shanker2004,kobidze2005}:
\begin{equation}
\label{eq:e_expansion}
\mathbf{E}(\mathbf{r}, t)=\sum_{n=1}^{N^{\mathrm{E}}}\{\mathbf{I}^{\mathrm{E}}(t)\}_{n} \mathbf{f}_{n}^{\mathrm{E}}(\mathbf{r})
\end{equation}
where $\mathbf{f}^{\mathrm{E}}_n(\mathbf{r})=\mathbf{f}^{+}_n(\mathbf{r})$ are the half SWG basis functions associated with triangular patches (there is one half SWG function for each patch that is located on the surface of $V$ and two for each ``internal'' patch), $\left\{\mathbf{I}^{\mathrm{E}}(t)\right\}_{n}=I_{n}^{\mathrm{E}}(t)$ are the unknown time-dependent expansion coefficients, and $N^{\mathrm{E}}=2 N^{\mathrm{D}}-N^{\mathrm{B}}$, where $N^{\mathrm{B}}$ is the number of patches located on the surface of $V$.

Inserting~\eqref{eq:d_expansion} and~\eqref{eq:e_expansion} into~\eqref{eq:td_efvie} and testing the resulting equation with $\mathbf{f}_{m}^{\mathrm{E}}(\mathbf{r})$, $m=1, \ldots, N^{\mathrm{E}}$ yield the spatially-discretized time-dependent TD-EFVIE as
\begin{equation}
\label{eq:semi_efvie}
\mathbf{G}^{\mathrm{EE}} \partial_{t} \mathbf{I}^{\mathrm{E}}(t)=\mathbf{V}^{\mathrm{inc}}(t)-\mathbf{V}^{\mathrm{sca}, \mathrm{E}}(t)-\mathbf{V}^{\mathrm{sca}, \mathrm{D}}(t).
\end{equation}
Here, the elements of the Gram matrix $\mathbf{G}^{\mathrm{EE}}$, the tested incident field vector $\mathbf{V}^{\mathrm{inc}}(t)$, and the tested scattered field vectors $\mathbf{V}^{\mathrm{sca}, \mathrm{E}}(t)$ and $\mathbf{V}^{\mathrm{sca}, \mathrm{D}}(t)$ are given by 
\begin{align}
\label{eq:gee} &\!\!\!\!\{\mathbf{G}^{\mathrm{EE}}\}_{m n}=\int_{V^{+}_{m}}\mathbf{f}_{m}^{\mathrm{E}}(\mathbf{r}) \cdot \mathbf{f}_{n}^{\mathrm{E}}(\mathbf{r}) d v \\
\label{eq:vinc}&\!\!\!\!\{\mathbf{V}^{\mathrm{inc}}(t)\}_{m}=\int_{V^{+}_{m}} \mathbf{f}_{m}^{\mathrm{E}}(\mathbf{r}) \cdot \partial_{t} \mathbf{E}^{\mathrm{inc}}(\mathbf{r}, t) d v \\
\label{eq:vscae}&\!\!\!\!\{\mathbf{V}^{\mathrm{sca}, \mathrm{E}}(t)\}_{m}=\sum_{n=1}^{N^{\mathrm{E}}} \int_{V^{+}_{m}} \mathbf{f}_{m}^{\mathrm{E}}(\mathbf{r}) \cdot \mathcal{L}[I_n^{\mathrm{E}}\mathbf{f}_{n}^{\mathrm{E}}](\mathbf{r},t)dv \\
\label{eq:vscad}&\!\!\!\!\{\mathbf{V}^{\mathrm{sca}, \mathrm{D}}(t)\}_{m}=-\frac{1}{\varepsilon_{0}}\!\sum_{n=1}^{N^{\mathrm{D}}} \int_{V^{+}_{m}}\! \mathbf{f}_{m}^{\mathrm{E}}(\mathbf{r}) \cdot \mathcal{L}[I_n^{\mathrm{D}}\mathbf{f}_{n}^{\mathrm{D}}](\mathbf{r},t)dv.
\end{align}

The semi-discretized system in~\eqref{eq:semi_efvie} is integrated in time using the $PE(CE)^m$ scheme described in Section~\ref{sec:pece}. This calls for sampling~\eqref{eq:semi_efvie} in time with time step size $\Delta t$. To compute the samples of $\mathbf{V}^{\mathrm{sca,E}}(t)$ and $\mathbf{V}^{\mathrm{sca,D}}(t)$, the retarded-time integrals $\mathcal{L}[I_n^{\mathrm{D}}\mathbf{f}_n^{\mathrm{D}}](\mathbf{r},t)$ and $\mathcal{L}[I_n^{\mathrm{E}}\mathbf{f}_n^{\mathrm{E}}](\mathbf{r},t)$ have to be evaluated at discrete times $t=j \Delta t$, which consequently means $\mathbf{I}^{\mathrm{E}}\left(t-R / c_{0}\right)$ and $\mathbf{I}^{\mathrm{D}}\left(t-R / c_{0}\right)$ (and their temporal derivatives) should be interpolated from the samples of $\mathbf{I}^{\mathrm{E}}(t)$ and $\mathbf{I}^{\mathrm{D}}(t)$, respectively. This requires expansion of $\mathbf{I}^{\mathrm{E}}(t)$ and $\mathbf{I}^{\mathrm{D}}(t)$ as
\begin{align}
\label{eq:temp_expansione}&\mathbf{I}^{\mathrm{E}}(t)=\sum_{i=1}^{N_{\mathrm{t}}} \mathbf{I}_{i}^{\mathrm{E}} T_{i}(t) \\
\label{eq:temp_expansiond}&\mathbf{I}^{\mathrm{D}}(t)=\sum_{i=1}^{N_{\mathrm{t}}} \mathbf{I}_{i}^{\mathrm{D}} T_{i}(t).
\end{align}
Here, $\mathbf{I}_{i}^{\mathrm{E}}=\mathbf{I}^{\mathrm{E}}(i \Delta t)$, $\mathbf{I}_{i}^{\mathrm{D}}=\mathbf{I}^{\mathrm{D}}(i \Delta t)$, $T_{i}(t)=T(t-i \Delta t)$, where $T(t)$ is the temporal interpolation function, and $N_{\mathrm{t}}$ is the number of time steps. Inserting~\eqref{eq:temp_expansione} and~\eqref{eq:temp_expansiond} into~\eqref{eq:semi_efvie} and point-testing the resulting equation in time, i.e., sampling it at $t=j \Delta t$, $j=1, \ldots, N_{\mathrm{t}}$ yield the fully discretized TD-EFVIE as 
\begin{equation}
\label{eq:full_efvie}
\mathbf{G}^{\mathrm{EE}} \dot{\mathbf{I}}_{j}^{\mathrm{E}}=\mathbf{V}_{j}^{\mathrm{inc}}-\mathbf{Z}_{0}^{\mathrm{EE}} \mathbf{I}_{j}^{\mathrm{E}}-\mathbf{Z}_{0}^{\mathrm{ED}} \mathbf{I}_{j}^{\mathrm{D}}
-\sum_{i=1}^{j-1} \mathbf{Z}_{j-i}^{\mathrm{EE}} \mathbf{I}_{i}^{\mathrm{E}}-\sum_{i=1}^{j-1} \mathbf{Z}_{j-i}^{\mathrm{ED}} \mathbf{I}_{i}^{\mathrm{D}}
\end{equation}
where $\dot{\mathbf{I}}_{j}^{\mathrm{E}}=\partial_{t} \mathbf{I}^{\mathrm{E}}(t)|_{t=j \Delta t}$, $\mathbf{V}_{j}^{\mathrm{inc}}=\mathbf{V}^{\mathrm{inc}}(j \Delta t)$, and the elements of the matrices $\mathbf{Z}_{j-i}^{\mathrm{EE}}$ and $\mathbf{Z}_{j-i}^{\mathrm{ED}}$ are given by
\begin{align}
\label{eq:zee}&\{\mathbf{Z}_{j-i}^{\mathrm{EE}}\}_{m n}=\int_{V^{+}_{m}} \mathbf{f}_{m}^{\mathrm{E}}(\mathbf{r}) \cdot \mathcal{L}[\mathbf{f}_{n}^{\mathrm{E}}T_{i}](\mathbf{r},j \Delta t)d v\\
\label{eq:zed}&\{\mathbf{Z}_{j-i}^{\mathrm{ED}}\}_{m n}=-\frac{1}{\varepsilon_{0}}\int_{V^{+}_{m}} \mathbf{f}_{m}^{\mathrm{E}}(\mathbf{r}) \cdot\mathcal{L}[\mathbf{f}_{n}^{\mathrm{D}}T_{i}](\mathbf{r},j \Delta t) d v.
\end{align}

One can see from the definition of the SWG function in~\eqref{eq:swg_def} that the full SWG functions can be expressed as linear combinations of the half SWG functions multiplied by the appropriate signs. In other words, basis set $\mathbf{f}_n^{\mathrm{D}}(\mathbf{r})$ can be constructed by linearly combining functions in basis set $\mathbf{f}_n^{\mathrm{E}}(\mathbf{r})$. This means that $\mathbf{Z}_{j-i}^{\mathrm{ED}}$ can be expressed in terms of $\mathbf{Z}_{j-i}^{\mathrm{EE}}$ using 
\begin{equation}
\label{eq:zed2}
\mathbf{Z}_{j-i}^{\mathrm{ED}}=-\frac{1}{\varepsilon_0}\mathbf{Z}_{j-i}^{\mathrm{EE}} \mathbf{P}^{\mathrm{T}}
\end{equation}
where $\mathbf{P}$ is the sparse matrix of linear mapping from the basis set $\mathbf{f}_n^{\mathrm{E}}(\mathbf{r})$ to the basis set $\mathbf{f}_n^{\mathrm{D}}(\mathbf{r})$, and its non-zero entries are either $1$ or $-1$. Inserting~\eqref{eq:zed2} into~\eqref{eq:full_efvie} yields the final form of the fully discretized TD-EFVIE as 
\begin{equation}
\label{eq:final_full_efvie}
\mathbf{G}^{\mathrm{EE}} \dot{\mathbf{I}}_{j}^{\mathrm{E}}=\mathbf{V}_{j}^{\mathrm{inc}}+\mathbf{Z}_{0}^{\mathrm{EE}}(\frac{1}{\varepsilon_0}\mathbf{P}^{\mathrm{T}}\mathbf{I}_{j}^{\mathrm{D}}-\mathbf{I}_{j}^{\mathrm{E}})
+\sum_{i=1}^{j-1} \mathbf{Z}_{j-i}^{\mathrm{EE}}(\frac{1}{\varepsilon_0}\mathbf{P}^{\mathrm{T}}\mathbf{I}_{i}^{\mathrm{D}}-\mathbf{I}_{i}^{\mathrm{E}}).
\end{equation}


The matrix entries $\{\mathbf{Z}_{j-i}^{\mathrm{EE}}\}_{m n}$ in~\eqref{eq:zee} can be explicitly written as
\begin{equation}
\label{eq:zee_exp}
\begin{aligned}
\{\mathbf{Z}_{j-i}^{\mathrm{EE}}\}_{m n}&=-\frac{\varepsilon_{0} \mu_{0}}{4 \pi} \int_{V^{+}_m} \mathbf{f}^{\mathrm{E}}_m(\mathbf{r})\cdot
\int_{V^{+}_n} \frac{1}{R} \mathbf{f}^{\mathrm{E}}_n(\mathbf{r}^{\prime})\partial_{t^{\prime}}^{3}T(t^{\prime}-i\Delta t)\big|_{t^{\prime}=j\Delta t-R / c_{0}} d v^{\prime}dv\\
+&\frac{1}{4 \pi} \int_{V^{+}_m}\mathbf{f}^{\mathrm{E}}_m(\mathbf{r})\cdot\nabla \int_{V^{+}_n} \frac{1}{R}\nabla^{\prime} \cdot\mathbf{f}^{\mathrm{E}}_n(\mathbf{r}^{\prime})\partial_{t^{\prime}} T(t^{\prime}-i\Delta t)\big|_{t^{\prime}=j\Delta t-R / c_{0}}d v^{\prime}dv.
\end{aligned}
\end{equation}
The order of the singularity in the second double integral in~\eqref{eq:zee_exp} is reduced by using the chain rule and the divergence theorem~\cite{gres2001, schaubert1984tetrahedral, sayed2015, zhang2017}, which yields
\begin{equation}
\label{eq:zee_exp2}
\begin{aligned}
\{\mathbf{Z}_{j-i}^{\mathrm{EE}}\}_{m n}&=-\frac{\varepsilon_{0} \mu_{0}}{4 \pi} \int_{V^{+}_m} \mathbf{f}^{\mathrm{E}}_m(\mathbf{r})\cdot
\int_{V^{+}_n} \frac{1}{R} \mathbf{f}^{\mathrm{E}}_n(\mathbf{r}^{\prime})\partial_{t^{\prime}}^{3}T(t^{\prime}-i\Delta t)\big|_{t^{\prime}=j\Delta t-R / c_{0}} d v^{\prime}dv\\
+&\frac{1}{4 \pi} \int_{S_m}\int_{V^{+}_n} \frac{1}{R}\nabla^{\prime} \cdot\mathbf{f}^{\mathrm{E}}_n(\mathbf{r}^{\prime})\partial_{t^{\prime}} T(t^{\prime}-i\Delta t)\big|_{t^{\prime}=j\Delta t-R / c_{0}}d v^{\prime}ds\\
-&\frac{1}{4 \pi} \int_{V^{+}_m}\nabla\cdot\mathbf{f}^{\mathrm{E}}_m(\mathbf{r})\int_{V^{+}_n} \frac{1}{R}\nabla^{\prime} \cdot\mathbf{f}^{\mathrm{E}}_n(\mathbf{r}^{\prime})\partial_{t^{\prime}} T(t^{\prime}-i\Delta t)\big|_{t^{\prime}=j\Delta t-R / c_{0}}d v^{\prime}dv.
\end{aligned}
\end{equation}
Note that the derivation of the surface integral expression in~\eqref{eq:zee_exp2} uses the fact that the normal component of $\mathbf{f}^{\mathrm{E}}_m(\mathbf{r})$ is equal to $1$ on $S_m$ and $0$ on the other three surfaces of $V_m^{+}$.

\subsection{Nonlinear Constitutive Relation}\label{sec:cons_relation}
In addition to TD-EFVIE in~\eqref{eq:td_efvie}, $\mathbf{E}(\mathbf{r}, t)$ and $\mathbf{D}(\mathbf{r}, t)$ for $\mathbf{r} \in V$ are related to each other via the nonlinear constitutive relation~\cite{saleh2007fundamentals, taflove2005computational}:
\begin{equation}
\label{eq:cons_relation}
\mathbf{D}(\mathbf{r}, t)=\varepsilon(\mathbf{E}) \mathbf{E}(\mathbf{r}, t), \quad \mathbf{r} \in V.
\end{equation}
Here, $\varepsilon(\mathbf{E})$ is the electric-field dependent permittivity and is expressed as
\begin{equation}
\label{eq:permittivity}
\varepsilon(\mathbf{E})=\varepsilon_{0}\left[\chi^{(1)}+\chi^{(3)}|\mathbf{E}(\mathbf{r}, t)|^{2}\right]
\end{equation}
where $\chi^{(1)}$ and $\chi^{(3)}$ are the linear and the third-order nonlinear coefficients associated with the Kerr nonlinearity, respectively.

The constitutive relation in~\eqref{eq:cons_relation} complements TD-EFVIE in~\eqref{eq:td_efvie}. Inserting~\eqref{eq:d_expansion} and~\eqref{eq:e_expansion} into~\eqref{eq:cons_relation} and testing the resulting equation with $\mathbf{f}_{m}^{\mathrm{D}}(\mathbf{r})$, $m=1, \ldots, N^{\mathrm{D}}$ at $t=j \Delta t$, $j=1, \ldots, N_{\mathrm{t}}$ yield
\begin{equation}
\label{eq:full_cons}
\mathbf{G}^{\mathrm{DD}} \mathbf{I}_{j}^{\mathrm{D}}=\mathbf{G}_{j}^{\mathrm{DE}} \mathbf{I}_{j}^{\mathrm{E}}.
\end{equation}
Here, the elements of the Gram matrices $\mathbf{G}^{\mathrm{DD}}$ and $\mathbf{G}_{j}^{\mathrm{DE}}$ are given by
\begin{align}
\label{eq:gdd}&\{\mathbf{G}^{\mathrm{DD}}\}_{m n}=\int_{V_{m}} \mathbf{f}_{m}^{\mathrm{D}}(\mathbf{r}) \cdot \mathbf{f}_{n}^{\mathrm{D}}(\mathbf{r}) d v \\
\label{eq:gde}&\{\mathbf{G}_{j}^{\mathrm{DE}}\}_{m n}=\int_{V_{m}} \mathbf{f}_{m}^{\mathrm{D}}(\mathbf{r}) \cdot \varepsilon(\mathbf{E}(\mathbf{r},j\Delta t)) \mathbf{f}_{n}^{\mathrm{E}}(\mathbf{r})d v
\end{align}
where $V_m = V_m^+ \cup V_m^-$ is the support of $\mathbf{f}_m^{\mathrm{D}}(\mathbf{r})$. Using $\mathbf{P}$, the sparse matrix of linear mapping from the basis set $\mathbf{f}_n^{\mathrm{E}}(\mathbf{r})$ to the basis set $\mathbf{f}_n^{\mathrm{D}}(\mathbf{r})$, one can express $\mathbf{G}^{\mathrm{DD}}$ in terms of $\mathbf{G}^{\mathrm{EE}}$ using
\begin{equation}
\label{eq:gdd2}
\mathbf{G}^{\mathrm{DD}}=\mathbf{P G}^{\mathrm{EE}} \mathbf{P}^{\mathrm{T}}.
\end{equation}
It is assumed that $\varepsilon(\mathbf{E})$ is a piece-wise constant function inside the scatterer, with a constant value in each tetrahedron. This constant value is computed at the center of each tetrahedron. Let $\mathbf{r}^{\mathrm c}_n$ represent the center of $V_n^{\pm}$. Inserting~\eqref{eq:e_expansion} into~\eqref{eq:permittivity} and evaluating the resulting expression at $\mathbf{r}=\mathbf{r}^{\mathrm{c}}_n$ and $t=j\Delta t$ yield 
\begin{equation}
\label{eq:permittivity_center}
\varepsilon(\mathbf{E}(\mathbf{r}_n^{\mathrm{c}},j\Delta t))=\varepsilon_{0}\bigg(\chi^{(1)}+\chi^{(3)}\bigg|\sum_{l}\left\{\mathbf{I}_{j}^{\mathrm{E}}\right\}_{l} \mathbf{f}_{l}^{\mathrm{E}}(\mathbf{r}_{n}^{\mathrm{c}})\bigg|^{2}\bigg)
\end{equation}
where the index $l$ runs over the indices of the half basis functions defined in $V_n^{+}$ (there are four of them). Let $\mathbf{S}^{\mathrm{E}}_j$ represent a diagonal matrix with entries 
\begin{equation}
\label{eq:se_j}
\{\mathbf{S}^{\mathrm{E}}_j\}_{nn}=\varepsilon(\mathbf{E}(\mathbf{r}_n^{\mathrm{c}},j\Delta t))=\varepsilon_{0}\bigg(\chi^{(1)}+\chi^{(3)}\bigg|\sum_{l}\left\{\mathbf{I}_{j}^{\mathrm{E}}\right\}_{l} \mathbf{f}_{l}^{\mathrm{E}}(\mathbf{r}_{n}^{\mathrm{c}})\bigg|^{2}\bigg).
\end{equation}
Then, $\mathbf{G}_{j}^{\mathrm{DE}}$ can be expressed in terms of $\mathbf{G}^{\mathrm{EE}}$ as
\begin{equation}
\label{eq:gde2}
\mathbf{G}_{j}^{\mathrm{DE}}=\mathbf{P}\mathbf{G}^{\mathrm{EE}} \mathbf{S}^{\mathrm{E}}_{j}.
\end{equation}
Inserting~\eqref{eq:gdd2} and~\eqref{eq:gde2} into~\eqref{eq:full_cons} yields the final form of the discretized constitutive relation as
\begin{equation}
\label{eq:final_full_cons}
\mathbf{P} \mathbf{G}^{\mathrm{EE}} \mathbf{P}^{\mathrm{T}} \mathbf{I}_{j}^{\mathrm{D}}=\mathbf{P} \mathbf{G}^{\mathrm{EE}} \mathbf{S}^{\mathrm{E}}_{j} \mathbf{I}_{j}^{\mathrm{E}}.
\end{equation}

The $PE(CE)^m$ scheme described in Section~\ref{sec:pece} requires $\mathbf{I}^{\mathrm{E}}_j$ to be updated from $\mathbf{I}^{\mathrm{D}}_j$ (during evaluation steps). This calls for the ``inversion'' of the nonlinear constitutive relation in~\eqref{eq:cons_relation}. This is done using the Padé approximant~\cite{oskooi2010Meep,baker1996,pozzi1994}: 
\begin{equation}
\label{eq:pade_approx}
\!\!\!\mathbf{E}(\mathbf{r},t)=\underbrace{\frac{1}{\varepsilon_{0} \chi^{(1)}}\left[\frac{\varepsilon_{0}^2(\chi^{(1)})^{3}+2 \chi^{(3)}|\mathbf{D}(\mathbf{r},t)|^{2}}{\varepsilon_{0}^2(\chi^{(1)})^{3}+3 \chi^{(3)}|\mathbf{D}(\mathbf{r},t)|^{2}}\right]}_{\displaystyle \tilde{\varepsilon}(\mathbf{D})}{\mathbf{D}(\mathbf{r},t)}.\!\!\!
\end{equation}
 Inserting~\eqref{eq:d_expansion} and~\eqref{eq:e_expansion} into~\eqref{eq:pade_approx} and testing the resulting equation with $\mathbf{f}_{m}^{\mathrm{E}}(\mathbf{r})$, $m=1, \ldots, N^{\mathrm{E}}$ at $t=j \Delta t$, $j=1, \ldots, N_{\mathrm{t}}$ yield
\begin{equation}
\label{eq:full_pade}
\mathbf{G}^{\mathrm{EE}} \mathbf{I}_{j}^{\mathrm{E}}=\mathbf{G}_{j}^{\mathrm{ED}} \mathbf{I}_{j}^{\mathrm{D}}.
\end{equation}
Here, the elements of the Gram matrix $\mathbf{G}_{j}^{\mathrm{ED}}$ are given by
\begin{equation}
\label{eq:ged}
\{\mathbf{G}_{j}^{\mathrm{ED}}\}_{m n}=\int_{V_{m}^{+}} \mathbf{f}_{m}^{\mathrm{E}}(\mathbf{r}) \cdot\tilde{\varepsilon}(\mathbf{D}(\mathbf{r},j\Delta t)) \mathbf{f}_{n}^{\mathrm{D}}(\mathbf{r})d v.
\end{equation}
Just like $\varepsilon(\mathbf{E})$, it is assumed that $\tilde{\varepsilon}(\mathbf{D})$ in~\eqref{eq:pade_approx} is a piece-wise constant function inside the scatterer, with a constant value in each tetrahedron. Inserting~\eqref{eq:d_expansion} into the expression of $\tilde{\varepsilon}(\mathbf{D})$ [see~\eqref{eq:pade_approx}] and evaluating the resulting expression at $\mathbf{r}=\mathbf{r}^{\mathrm{c}}_n$ and $t=j\Delta t$ yield 
\begin{equation}
\label{eq:inverse_permittivity_center}
\tilde{\varepsilon}(\mathbf{D}(\mathbf{r}_n^{\mathrm{c}},j\Delta t))=\frac{1}{\varepsilon_{0} \chi^{(1)}}\frac{\displaystyle \bigg(\varepsilon_{0}^2(\chi^{(1)})^3+2\chi^{(3)}\bigg|\sum_{l}\left\{\mathbf{I}_{j}^{\mathrm{D}}\right\}_{l} \mathbf{f}_{l}^{\mathrm{D}}(\mathbf{r}_n^{\mathrm{c}})\bigg|^{2}\bigg)}{\displaystyle \bigg(\varepsilon_{0}^2(\chi^{(1)})^3+3\chi^{(3)}\bigg|\sum_{l}\left\{\mathbf{I}_{j}^{\mathrm{D}}\right\}_{l} \mathbf{f}_{l}^{\mathrm{D}}(\mathbf{r}_n^{\mathrm{c}})\bigg|^{2}\bigg)}
\end{equation}
where the index $l$ runs over the indices of the basis functions that have $V_n^{+}$ or $V_n^{-}$ as support. Let $\mathbf{S}^{\mathrm{D}}_j$ represent a diagonal matrix with entries
\begin{equation}
\label{eq:sd_j}
\{\mathbf{S}^{\mathrm{D}}_j\}_{nn}=\tilde{\varepsilon}(\mathbf{D}(\mathbf{r}_n^{\mathrm{c}},j\Delta t))=\frac{1}{\varepsilon_{0} \chi^{(1)}}\frac{\displaystyle \bigg(\varepsilon_{0}^2(\chi^{(1)})^3+2\chi^{(3)}\bigg|\sum_{l}\left\{\mathbf{I}_{j}^{\mathrm{D}}\right\}_{l} \mathbf{f}_{l}^{\mathrm{D}}(\mathbf{r}_n^{\mathrm{c}})\bigg|^{2}\bigg)}{\displaystyle \bigg(\varepsilon_{0}^2(\chi^{(1)})^3+3\chi^{(3)}\bigg|\sum_{l}\left\{\mathbf{I}_{j}^{\mathrm{D}}\right\}_{l} \mathbf{f}_{l}^{\mathrm{D}}(\mathbf{r}_n^{\mathrm{c}})\bigg|^{2}\bigg)}.
\end{equation}
Then, $\mathbf{G}_{j}^{\mathrm{ED}}$ can be expressed in terms of $\mathbf{G}^{\mathrm{EE}}$ using
\begin{equation}
\label{eq:ged2}
\mathbf{G}_{j}^{\mathrm{ED}}=\mathbf{G}^{\mathrm{EE}}\mathbf{S}^{\mathrm{D}}_{j}\mathbf{P}^{\mathrm{T}}.
\end{equation}
Inserting~\eqref{eq:ged2} into~\eqref{eq:full_pade} and eliminating $\mathbf{G}^{\mathrm{EE}}$ from both sides of the resulting equation yield the final form of the discretized Padé approximant as:
\begin{equation}
\label{eq:final_full_pade}
\mathbf{I}_{j}^{\mathrm{E}}=\mathbf{S}^{\mathrm{D}}_{j}\mathbf{P}^{\mathrm{T}} \mathbf{I}_{j}^{\mathrm{D}}.
\end{equation}

\subsection{PE(CE)m Scheme}\label{sec:pece}

The fully-discretized TD-EFVIE~\eqref{eq:final_full_efvie} relates unknowns $\mathbf{I}_{j}^{\mathrm{E}}$ and $\mathbf{I}_{j}^{\mathrm{D}}$ to the time derivative of the unknown $\dot{\mathbf{I}}_{j}^{\mathrm{E}}$, and is integrated in time using a $PE(CE)^m$ scheme to yield the unknown $\mathbf{I}_{j}^{\mathrm{E}}$~\cite{ulku2013,chen2019,sayed2020,chen2021}. This scheme uses the discretized constitutive relation~\eqref{eq:final_full_cons} and the discretized Padé approximant~\eqref{eq:final_full_pade} to update $\mathbf{I}_{j}^{\mathrm{E}}$ and $\mathbf{I}_{j}^{\mathrm{D}}$. The steps of the $PE(CE)^m$ are provided as follows.\\

\noindent \emph{Loop over} $j=1, \ldots, N_{\mathrm{t}}$.\\

\noindent  \emph{Step 0:} Compute $\mathbf{V}_{j}^{\mathrm{fix }}$, the part of the right-hand side of~\eqref{eq:final_full_efvie} that does not change within the time step $j$:
\begin{equation}
\label{eq:step0}
\mathbf{V}_{j}^{\mathrm{fix}}=\mathbf{V}_{j}^{\mathrm{inc}}+\sum_{i=1}^{j-1} \mathbf{Z}_{j-i}^{\mathrm{EE}}(\frac{1}{\varepsilon_0}\mathbf{P}^{\mathrm{T}}\mathbf{I}_{i}^{\mathrm{D}}-\mathbf{I}_{i}^{\mathrm{E}}).
\end{equation}

\noindent \underline{$PE$ stage}\\

\noindent  \emph{Step 1:} Predict $\mathbf{I}_{j}^{\mathrm{E},(0)}$ using $\mathbf{I}_{i}^{\mathrm{E}}$ and $\dot{\mathbf{I}}_{i}^{\mathrm{E}}$, $i=j-k, \ldots, j-1$:
\begin{equation}
\label{eq:step1}
\mathbf{I}_{j}^{\mathrm{E},(0)}=\sum_{l=1}^{k}\left[\{\mathbf{p}\}_{l} \mathbf{I}_{j-1+l-k}^{\mathrm{E}}+\{\mathbf{p}\}_{k+l} \dot{\mathbf{I}}_{j-1+l-k}^{\mathrm{E}}\right].
\end{equation}

\noindent \emph{Step 2:} Compute $\mathbf{S}_j^{\mathrm{E},(0)}$ by inserting $\mathbf{I}_{j}^{\mathrm{E},(0)}$ into~\eqref{eq:se_j}:
\begin{equation}
\label{eq:step2}
\{\mathbf{S}^{\mathrm{E},(0)}_j\}_{nn}=\\
\varepsilon_{0}\bigg(\chi^{(1)}+\chi^{(3)}\bigg|\sum_{l}\left\{\mathbf{I}_{j}^{\mathrm{E},(0)}\right\}_{l} \mathbf{f}_{l}^{\mathrm{E}}(\mathbf{r}_{n}^{\mathrm{c}})\bigg|^{2}\bigg).
\end{equation}
\noindent \emph{Step 3:} Compute $\mathbf{I}_{j}^{\mathrm{D},(0)}$ by solving~\eqref{eq:final_full_cons} with $\mathbf{I}_{j}^{\mathrm{E},(0)}$ and $\mathbf{S}_j^{\mathrm{E},(0)}$:
\begin{equation}
\label{eq:step3}
\mathbf{P} \mathbf{G}^{\mathrm{EE}} \mathbf{P}^{\mathrm{T}} \mathbf{I}_{j}^{\mathrm{D},(0)}=\mathbf{P} \mathbf{G}^{\mathrm{EE}} \mathbf{S}^{\mathrm{E},(0)}_{j} \mathbf{I}_{j}^{\mathrm{E},(0)}.
\end{equation}
\noindent  \emph{Step 4:} Compute $\mathbf{S}_j^{\mathrm{D},(0)}$ by inserting $\mathbf{I}_{j}^{\mathrm{D},(0)}$ into~\eqref{eq:sd_j}: 
\begin{equation}
\label{eq:step4}
\{\mathbf{S}^{\mathrm{D},(0)}_j\}_{nn}=\frac{1}{\varepsilon_{0} \chi^{(1)}}\frac{\displaystyle \bigg(\varepsilon_{0}^2(\chi^{(1)})^3+2\chi^{(3)}\bigg|\sum_{l}\left\{\mathbf{I}_{j}^{\mathrm{D},(0)}\right\}_{l} \mathbf{f}_{l}^{\mathrm{D}}(\mathbf{r}_n^{\mathrm{c}})\bigg|^{2}\bigg)}{\displaystyle \bigg(\varepsilon_{0}^2(\chi^{(1)})^3+3\chi^{(3)}\bigg|\sum_{l}\left\{\mathbf{I}_{j}^{\mathrm{D},(0)}\right\}_{l} \mathbf{f}_{l}^{\mathrm{D}}(\mathbf{r}_n^{\mathrm{c}})\bigg|^{2}\bigg)}.
\end{equation}
\noindent  \emph{Step 5:} Update $\mathbf{I}_{j}^{\mathrm{E},(0)}$ by using $\mathbf{I}_{j}^{\mathrm{D},(0)}$ and $\mathbf{S}_j^{\mathrm{D},(0)}$ in~\eqref{eq:final_full_pade}:
\begin{equation}
\label{eq:step5}
\mathbf{I}_{j}^{\mathrm{E},(0)}=\mathbf{S}^{\mathrm{D},(0)}_{j}\mathbf{P}^{\mathrm{T}} \mathbf{I}_{j}^{\mathrm{D},(0)}.
\end{equation}

\noindent  \emph{Step 6:} Compute $\dot{\mathbf{I}}_{j}^{\mathrm{E},(0)}$ by solving
\begin{equation}
\label{eq:step6}
\mathbf{G}^{\mathrm{EE}} \dot{\mathbf{I}}_{j}^{\mathrm{E},(0)}=\mathbf{V}_{j}^{\mathrm{fix}}+\mathbf{Z}_{0}^{\mathrm{EE}}\left(\frac{1}{\varepsilon_{0}} \mathbf{P}^{\mathrm{T}} \mathbf{I}_{j}^{\mathrm{D},(0)}-\mathbf{I}_{j}^{\mathrm{E},(0)}\right).
\end{equation}

\noindent \underline{$(CE)^m$ stage}\\

\noindent  \emph{Step 7:} \emph{Loop over} $m=1, \ldots, m_{\mathrm{max}}$:\\

\emph{Step 7.1:} Correct $\mathbf{I}_{j}^{\mathrm{E},(m)}$ using $\mathbf{I}_{i}^{\mathrm{E}}$ and $\dot{\mathbf{I}}_{i}^{\mathrm{E}}$, $i=j-k, \ldots, j-1$, and $\dot{\mathbf{I}}_{j}^{\mathrm{E},(m-1)}$:
\begin{equation}
\label{eq:step7_1}
\mathbf{I}_{j}^{\mathrm{E},(m)}=\sum_{l=1}^{k}\left[\{\mathbf{c}\}_{l} \mathbf{I}_{j-1+l-k}^{\mathrm{E}}+\{\mathbf{c}\}_{k+l} \dot{\mathbf{I}}_{j-1+l-k}^{\mathrm{E}}\right]+\{\mathbf{c}\}_{2k+1} \dot{\mathbf{I}}_{j}^{\mathrm{E},(m-1)}.
\end{equation}

\emph{Step 7.2:} Apply successive over relaxation (SOR) to $\mathbf{I}_{j}^{\mathrm{E},(m)}$ with $\alpha \in[0,1]$:
\begin{equation}
\label{eq:step7_2}
\mathbf{I}_{j}^{\mathrm{E},(m)}=\alpha \mathbf{I}_{j}^{\mathrm{E},(m)}+(1-\alpha) \mathbf{I}_{j}^{\mathrm{E},(m-1)}.
\end{equation}

\emph{Step 7.3:} Compute $\mathbf{S}_j^{\mathrm{E},(m)}$ by inserting $\mathbf{I}_{j}^{\mathrm{E},(m)}$ into~\eqref{eq:se_j}:
\begin{equation}
\label{eq:step7_3}
\{\mathbf{S}^{\mathrm{E},(m)}_j\}_{nn}=\\
\varepsilon_{0}\bigg(\chi^{(1)}+\chi^{(3)}\bigg|\sum_{l}\left\{\mathbf{I}_{j}^{\mathrm{E},(m)}\right\}_{l} \mathbf{f}_{l}^{\mathrm{E}}(\mathbf{r}_{n}^{\mathrm{c}})\bigg|^{2}\bigg).
\end{equation}

\emph{Step 7.4:} Compute $\mathbf{I}_{j}^{\mathrm{D},(m)}$ by solving~\eqref{eq:final_full_cons} with $\mathbf{I}_{j}^{\mathrm{E},(m)}$ and $\mathbf{S}_j^{\mathrm{E},(m)}$:
\begin{equation}
\label{eq:step7_4}
\mathbf{P} \mathbf{G}^{\mathrm{EE}} \mathbf{P}^{\mathrm{T}} \mathbf{I}_{j}^{\mathrm{D},(m)}=\mathbf{P} \mathbf{G}^{\mathrm{EE}} \mathbf{S}_{j}^{\mathrm{E},(m)} \mathbf{I}_{j}^{\mathrm{E},(m)}.
\end{equation}

\emph{Step 7.5:} Compute $\mathbf{S}_j^{\mathrm{D},(m)}$ by inserting $\mathbf{I}_{j}^{\mathrm{D},(m)}$ into~\eqref{eq:sd_j}: 
\begin{equation}
\label{eq:step7_5}
\{\mathbf{S}^{\mathrm{D},(m)}_j\}_{nn}=\frac{1}{\varepsilon_{0} \chi^{(1)}}
\frac{\displaystyle \bigg(\varepsilon_{0}^2(\chi^{(1)})^3+2\chi^{(3)}\bigg|\sum_{l}\left\{\mathbf{I}_{j}^{\mathrm{D},(m)}\right\}_{l} \mathbf{f}_{l}^{\mathrm{D}}(\mathbf{r}_n^{\mathrm{c}})\bigg|^{2}\bigg)}{\displaystyle \bigg(\varepsilon_{0}^2(\chi^{(1)})^3+3\chi^{(3)}\bigg|\sum_{l}\left\{\mathbf{I}_{j}^{\mathrm{D},(m)}\right\}_{l} \mathbf{f}_{l}^{\mathrm{D}}(\mathbf{r}_n^{\mathrm{c}})\bigg|^{2}\bigg)}.
\end{equation}

\emph{Step 7.6:} Update $\mathbf{I}_{j}^{\mathrm{E},(m)}$ by using $\mathbf{I}_{j}^{\mathrm{D},(m)}$ and $\mathbf{S}_j^{\mathrm{D},(m)}$ in~\eqref{eq:final_full_pade}:
\begin{equation}
\label{eq:step7_6}
\mathbf{I}_{j}^{\mathrm{E},(m)}=\mathbf{S}^{\mathrm{D},(m)}_{j}\mathbf{P}^{\mathrm{T}} \mathbf{I}_{j}^{\mathrm{D},(m)}.
\end{equation}

\emph{Step 7.7:} Compute $\dot{\mathbf{I}}_{j}^{\mathrm{E},(m)}$ by solving
\begin{equation}
\label{eq:step7_7}
\mathbf{G}^{\mathrm{EE}} \dot{\mathbf{I}}_{j}^{\mathrm{E},(m)}=\mathbf{V}_{j}^{\mathrm{fix}}+\mathbf{Z}_{0}^{\mathrm{EE}}\left(\frac{1}{\varepsilon_{0}} \mathbf{P}^{\mathrm{T}} \mathbf{I}_{j}^{\mathrm{D},(m)}-\mathbf{I}_{j}^{\mathrm{E},(m)}\right).
\end{equation}

\emph{Step 7.8:} Check convergence 
\begin{equation}
\label{eq:step7_8}
\left\|\mathbf{I}_{j}^{\mathrm{E},(m)}-\mathbf{I}_{j}^{\mathrm{E},(m-1)}\right\|<\left\|\mathbf{I}_{j}^{\mathrm{E},(m)}\right\| \epsilon^{\mathrm{PECE}}
\end{equation}
where $\epsilon^{\mathrm{PECE}}$ is the convergence threshold, and $\left\|\mathbf{x}\right\|$ represents the $L_2$-norm of vector $\mathbf{x}$. \\
\emph{End loop over} $m$.\\

\noindent \emph{Step 8:} Upon converge, set $\mathbf{I}_{j}^{\mathrm{E}}=\mathbf{I}_{j}^{\mathrm{E},(m)}, \mathbf{I}_{j}^{\mathrm{D}}=\mathbf{I}_{j}^{\mathrm{D},(m)}$, and $\dot{\mathbf{I}}_{j}^{\mathrm{E}}=\dot{\mathbf{I}}_{j}^{\mathrm{E},(m)}$.\\

\noindent \emph{End loop over} $j$. \\

\noindent In the $PE(CE)^m$ scheme described above, $\mathbf{p}$ and $\mathbf{c}$ are the predictor and corrector coefficient vectors of length $2 k$ and $2 k+1$, respectively~\cite{hairer2010,glaser2009new}. SOR in~\eqref{eq:step7_2} [\emph{Step 7.2}] helps to maintain the stability of the solution~\cite{sayed2014stable,liu2016}. The Gram matrix $\mathbf{G}^{\mathrm{EE}}$ is well-conditioned and sparse, and therefore the matrix systems in~\eqref{eq:step3} [\emph{Step 3}],~\eqref{eq:step6} [\emph{Step 6}],~\eqref{eq:step7_4} [\emph{Step 7.4}], and~\eqref{eq:step7_7} [\emph{Step 7.7}] are solved efficiently using a linear iterative solver. The iterations of this solver are terminated when the following convergence criterion is satisfied:
\begin{equation}
\label{eq:convergence}
\left\|\mathbf{I}_{j}^{(n)}-\mathbf{I}_{j}^{(n-1)}\right\|<\epsilon^{\mathrm{ITS}}\left\|\mathbf{b}\right\|.
\end{equation}
Here, $\epsilon^{\mathrm{ITS}}$ is the convergence threshold, $\mathbf{b}$ is the right-hand side vector, and $\mathbf{I}_{j}^{(n)}$ and $\mathbf{I}_{j}^{(n-1)}$ are the solutions at iterations $n$ and $n-1$, respectively.


\subsection{Comments}\label{sec:comments}
Several comments about the formulation and the discretization schemes described in Sections~\ref{sec:tdefvie} and~\ref{sec:cons_relation}, and the $PE(CE)^{m}$ scheme described in Section~\ref{sec:pece} are in order: 

1) The second-order nonlinear term (with coefficient $\chi^{(2)}$) is not considered in the expression of $\varepsilon(\mathbf{E})$ given by~\eqref{eq:permittivity} because it is assumed that the scatterer is centrosymmetric~\cite{saleh2007fundamentals}. But the method developed in this work could still be used for permittivity functions with second- and higher-order terms. 

2) The most common choices for the temporal interpolation function $T_i(t)$ used in~\eqref{eq:temp_expansione} and~\eqref{eq:temp_expansiond} are the Lagrange polynomials~\cite{manara1997,aygun2002,bagci2005} and the band limited approximate prolate spheroidal wave (APSW) functions~\cite{knab1979,weile2004,wildman2004,sayed2015}. Both options can be used by the MOT solver described in this work. However, when APSW functions are used, the resulting $PE(CE)^m$ scheme is no longer casual, i.e., ``future'' samples of $\mathbf{I}_{i}^{\mathrm{E}}$ and $\mathbf{I}_{i}^{\mathrm{D}}$ are required to compute the summation on the right-hand side of~\eqref{eq:step0}. In this case, the causality of the time marching is restored using the extrapolation scheme described in~\cite{sayed2015}. This extrapolation scheme is specifically tailored for the accurate and stable solution of TD-EFVIE. 

3) Similarly, several options exist to generate the predictor and corrector coefficients, $\mathbf{p}$ and $\mathbf{c}$. They can be obtained using polynomial interpolation, which leads to well-known linear multistep methods such as Adams–Moulton, Adams–Bashforth, and backward difference schemes~\cite{hairer2010}. One can also obtain them numerically under the assumption that oscillating and/or decaying exponential functions can be used to approximate the time-dependence of the solution~\cite{glaser2009new}.

4) When MOT-based TD-EFVIE solvers are used to analyze electromagnetic scattering from linear objects, $\Delta t$ is selected using $\Delta t=1 /\left(2 \gamma f_{\max }\right)$ where $\gamma$ is the oversampling factor. Ideally, $\gamma$ can be set to $1.0$ due to the Nyquist sampling criterion, but, in general, it is set to a value between $2.5$ and $15.0$ (depending on the desired level of accuracy). On the other hand, for nonlinear scatterers, there is no explicit guidance criterion to select $\Delta t$. For the numerical examples presented in Section~\ref{sec:numres}, $\gamma$ is increased to sufficiently resolve the frequency of the higher harmonics.

5) SOR in~\eqref{eq:step7_2} [\emph{Step 7.2}] balances between the stability and the convergence of the corrector updates $(CE)^m$, and it does not affect the accuracy of the solution (assuming convergence). Reducing the SOR coefficient $\alpha$ increases the number of corrector updates, which in return increases the computation time. On the other hand, increasing $\alpha$ might result in unstable corrector updates leading to an unstable solution. Additionally, for stronger nonlinearities, one must reduce $\alpha$ to maintain the stability, which again comes with increased computation time.

6) Within the framework of the proposed explicit MOT solver, a more accurate expansion of $\mathbf{E}(\mathbf{r},t)$ can be used. For example, the half SWG functions can be replaced by the fully linear curl-conforming basis functions as done in~\cite{sayed2020}. These functions are defined on the edges of the tetrahedrons and automatically enforce the tangential continuity of $\mathbf{E}(\mathbf{r},t)$, leading to a more accurate solution. Switching to this type of basis functions would not change how the nonlinearity is accounted for and the time integration is carried out using a $PE(CE)^m$ scheme during time marching. Having said that, using SWG and half SWG functions to expand $\mathbf{D}(\mathbf{r},t)$ and $\mathbf{E}(\mathbf{r},t)$, respectively, reduces the computational cost of the solver. This is because full SWG functions can be constructed by linearly combining half SWG functions leading to the sparse mappings between the matrices and the expansion coefficients as shown in~\eqref{eq:zed2},~\eqref{eq:gdd2},~\eqref{eq:gde2}, and~\eqref{eq:ged2}.

\begin{figure}[t!]
\centering
\subfigure[]{\includegraphics[width=0.49\columnwidth]{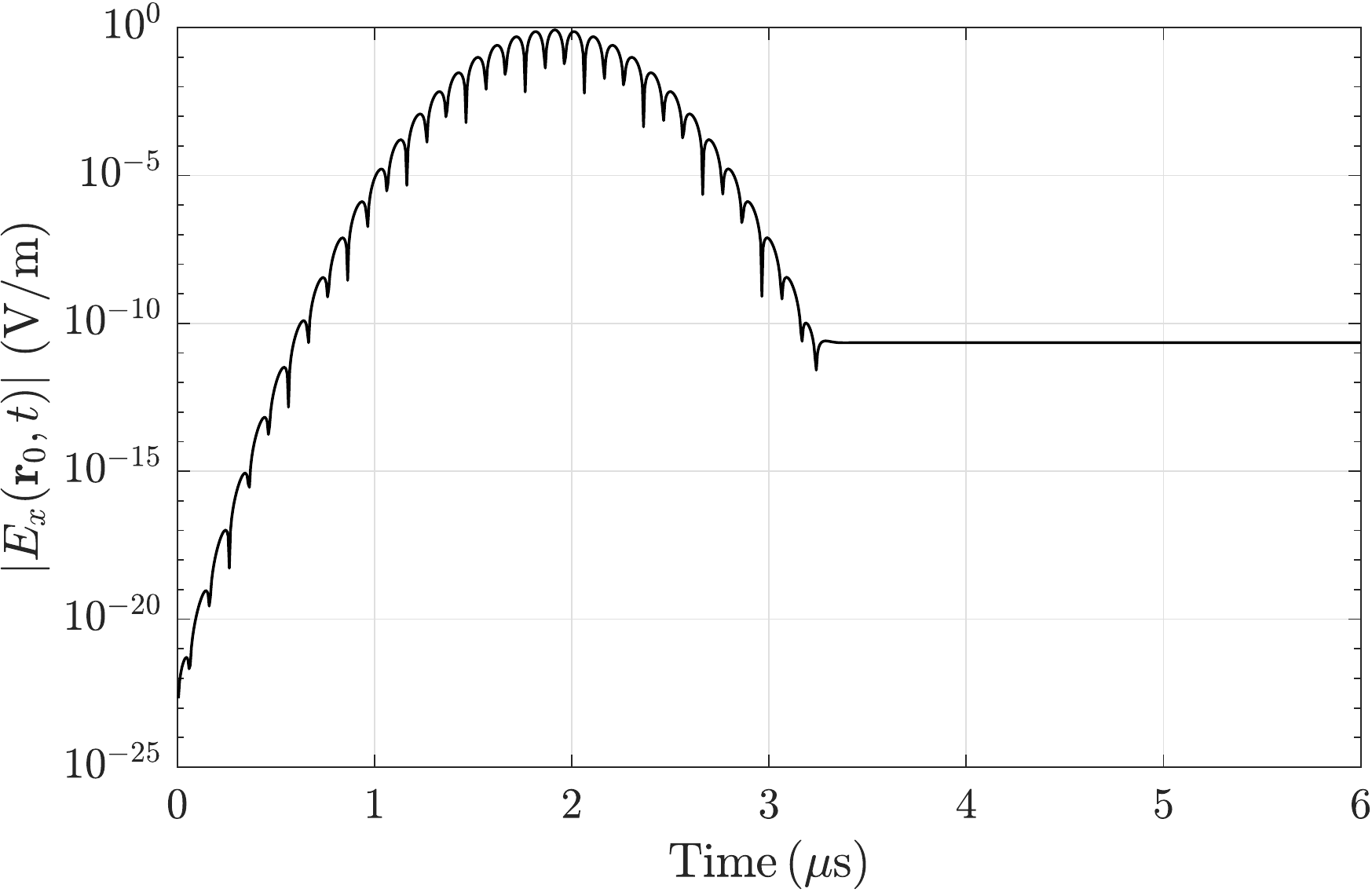}}
\subfigure[]{\includegraphics[width=0.49\columnwidth]{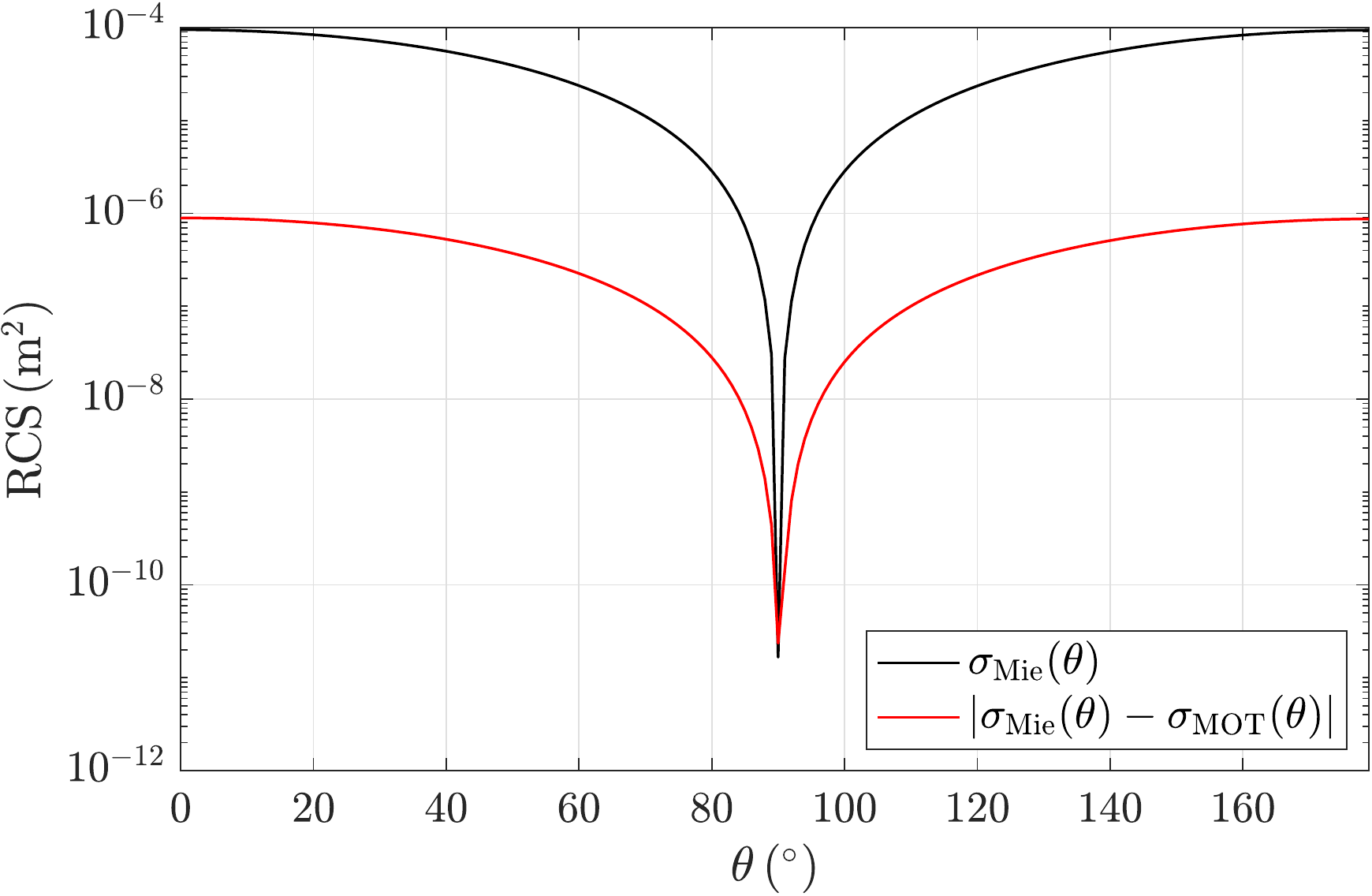}}
\caption{Scattering from a linear sphere. (a) $x$-component of $\mathbf{E}(\mathbf{r}, t)$ computed by the proposed solver at the center of the sphere [$\mathbf{r}_0=(0,0,0)$].
(b) $\sigma_{\mathrm{Mie}}(\theta)$ and $|\sigma_{\mathrm{Mie}}(\theta)-\sigma_{\mathrm{MOT}}(\theta)|$, where $\sigma_{\mathrm{MOT}}(\theta)$ and $\sigma_{\mathrm{Mie}}(\theta)$ are the RCS computed at $f=5.0\,\mathrm{MHz}$ on the $\phi=0$ plane using the Fourier-transformed time-domain solution and the Mie series solution, respectively.}
\label{fig:linear_sphere}
\end{figure}

\section{Numerical Results}\label{sec:numres}
This section presents several numerical examples that demonstrate the accuracy, stability, and applicability of the proposed explicit MOT solver in characterizing electromagnetic field interactions on scatterers with Kerr nonlinearity. In all the examples considered here, the scatterer resides in free space with permittivity $\varepsilon_0$ and permeability $\mu_0$. In all simulations, the excitation is a plane wave with electric field
\begin{equation}
\label{eq:excitation}
\mathbf{E}^{\mathrm{inc}}(\mathbf{r}, t)=\hat{\mathbf{p}} E_{0} P(t-\mathbf{r} \cdot \hat{\mathbf{k}} /c_{0})
\end{equation}
where $E_{0}$ is the amplitude of the electric field, and $\hat{\mathbf{p}}=\hat{\mathbf{x}}$ and $\hat{\mathbf{k}}=\hat{\mathbf{z}}$ are the unit vectors that represent the directions of the electric field and the propagation of the plane wave, respectively. In~\eqref{eq:excitation}, $P(t)$ is a band-limited pulse that describes the time dependence of the excitation. 

The predictor and corrector coefficient vectors, $\mathbf{p}$ and $\mathbf{c}$ are obtained using the sixth-order Adams-Bashforth and backward difference formulas~\cite{hairer2010}, respectively. The convergence threshold for the corrector updates [$(CE)^m$] is set to $\epsilon^{\mathrm{PECE}}=10^{-13}$ [see~\eqref{eq:step7_8}]. The matrix equations in~\eqref{eq:step3},~\eqref{eq:step6},~\eqref{eq:step7_4}, and~\eqref{eq:step7_7} are iteratively solved using the transpose-free quasi-minimal residual (TFQMR) method~\cite{freund1993transpose}. The convergence threshold of the TFQMR iterations is set to $\epsilon^{\mathrm{ITS}}=10^{-12}$ [see~\eqref{eq:convergence}].
\subsection{Linear Sphere}\label{sec:linear_sphere}

In the first example, electromagnetic scattering from a ``linear'' sphere is analyzed using the proposed method. The sphere is centered at the origin and has a radius of length $1.0\,\mathrm{m}$. The coefficients of the permittivity function of the sphere are $\chi^{(1)}=2.0$ and $\chi^{(3)}=0$. The time dependence of the plane wave excitation in~\eqref{eq:excitation} is a modulated Gaussian pulse expressed as
\begin{equation}
\label{eq:linear_sphere}
    P(t)=\cos(2 \pi f_0[t-t_{\mathrm{p}}])e^{-(t-t_{\mathrm{p}})^2/(2\sigma^2)}.
\end{equation}
Here, $f_0$, $t_{\mathrm{p}}$, and $\sigma$ are the modulation frequency, time delay, and duration of the pulse, respectively. Let $f_{\mathrm{bw}}$ represent the effective bandwidth, then choosing $\sigma=3 /\left(2 \pi f_{\mathrm{bw}}\right)$ ensures that $99.997\%$ of $P(t)$'s power is within the frequency band $[f_{\mathrm{min}},f_{\mathrm{max}}]$, where $f_{\mathrm{min}}=f_{0}-f_{\mathrm{bw}}$ and $f_{\mathrm{max}}=f_{0}+f_{\mathrm{bw}}$~\cite{bagci2007}. 

For this example, $f_{0}=5.0\,\mathrm{MHz}$, $f_{\mathrm{bw}}=2.5\,\mathrm{MHz}$, $\sigma =  0.1910\,\mu\mathrm{s}$, and $t_{\mathrm{p}}=8 \sigma$. $\mathbf{E}(\mathbf{r}, t)$ and $\mathbf{D}(\mathbf{r}, t)$ induced inside the sphere are discretized using $N^{\mathrm{E}}=3\,844$ and $N^{\mathrm{D}}=2\,114$ spatial basis functions, respectively. The MOT scheme is executed for $N_{\mathrm{t}}=1\,000$ time steps with $\Delta t=6.0\,\mathrm{ns}$. The SOR coefficient in~\eqref{eq:step7_2} is selected as $\alpha=0.3$.

Fig.~\ref{fig:linear_sphere}(a) plots the $x$-component of $\mathbf{E}(\mathbf{r}, t)$ computed by the proposed solver at the center of the sphere (the sphere is centered at the origin). The figure shows that the proposed solver provides stable results for the entire simulation duration. After the time domain simulation is completed, the Fourier-transformed solution is used to compute the radar cross section (RCS) $\sigma_{\mathrm{MOT}}(\theta)$ at $f=5.0\,\mathrm{MHz}$ on the $\phi=0$ plane ($\theta \in \left[0^{\circ},180^{\circ}\right]$). Let $\sigma_{\mathrm{Mie}}(\theta)$ represent the RCS computed at the same frequency and on the same plane using the Mie series solution~\cite{jin2011theory}. Fig.~\ref{fig:linear_sphere}(b) plots $\sigma_{\mathrm{Mie}}(\theta)$ and $|\sigma_{\mathrm{Mie}}(\theta)-\sigma_{\mathrm{MOT}}(\theta)|$ versus $\theta$. The figure clearly shows that the result obtained using the proposed solver is accurate.

\subsection{Nonlinear Cube}\label{sec:nonlinear_cube}
\begin{figure*}[t!]
\centering
\subfigure[]{\includegraphics[width=0.48\columnwidth]{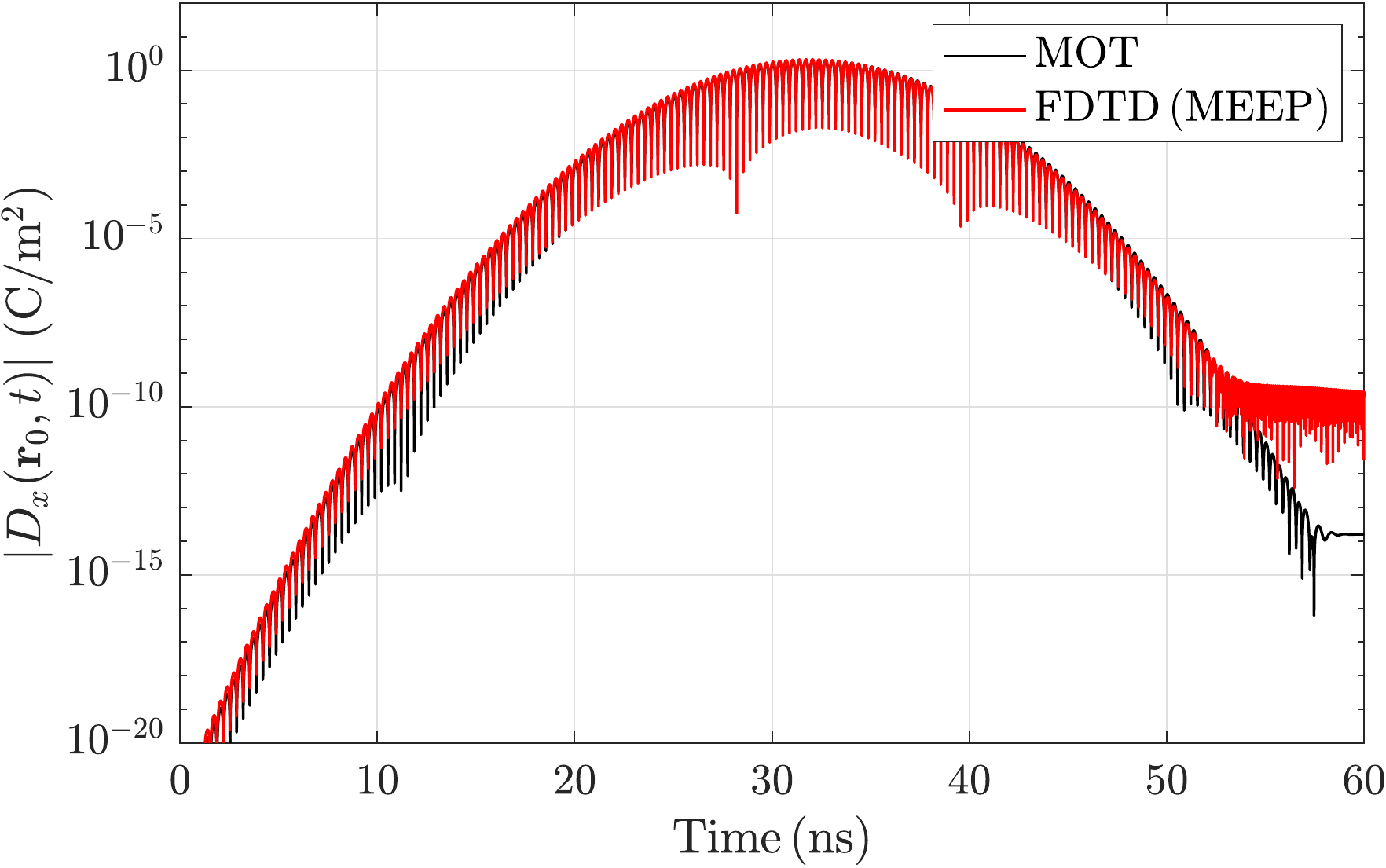}}\hspace{0.25cm}
\subfigure[]{\includegraphics[width=0.46\columnwidth]{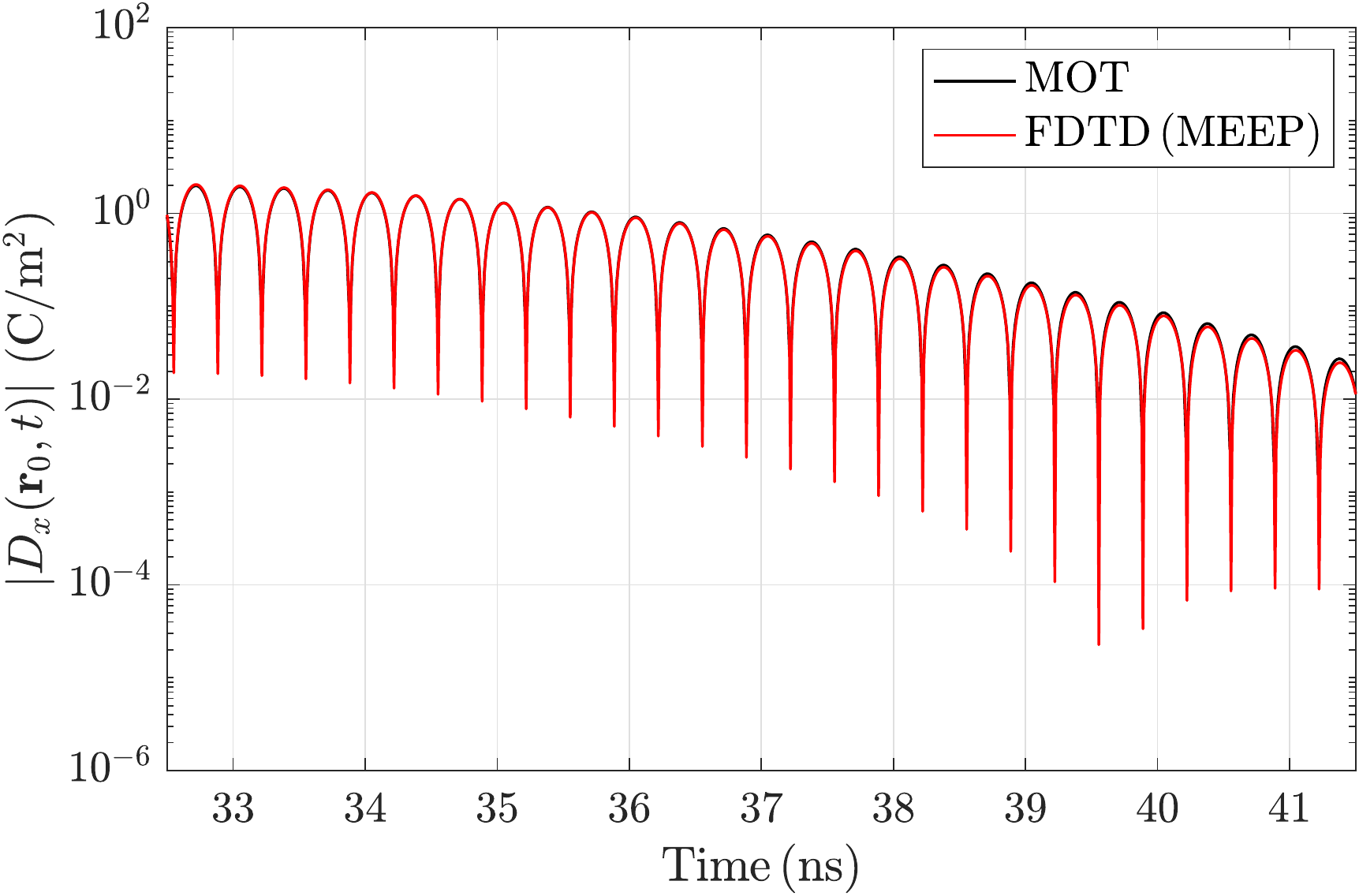}}
\subfigure[]{\includegraphics[width=0.48\columnwidth]{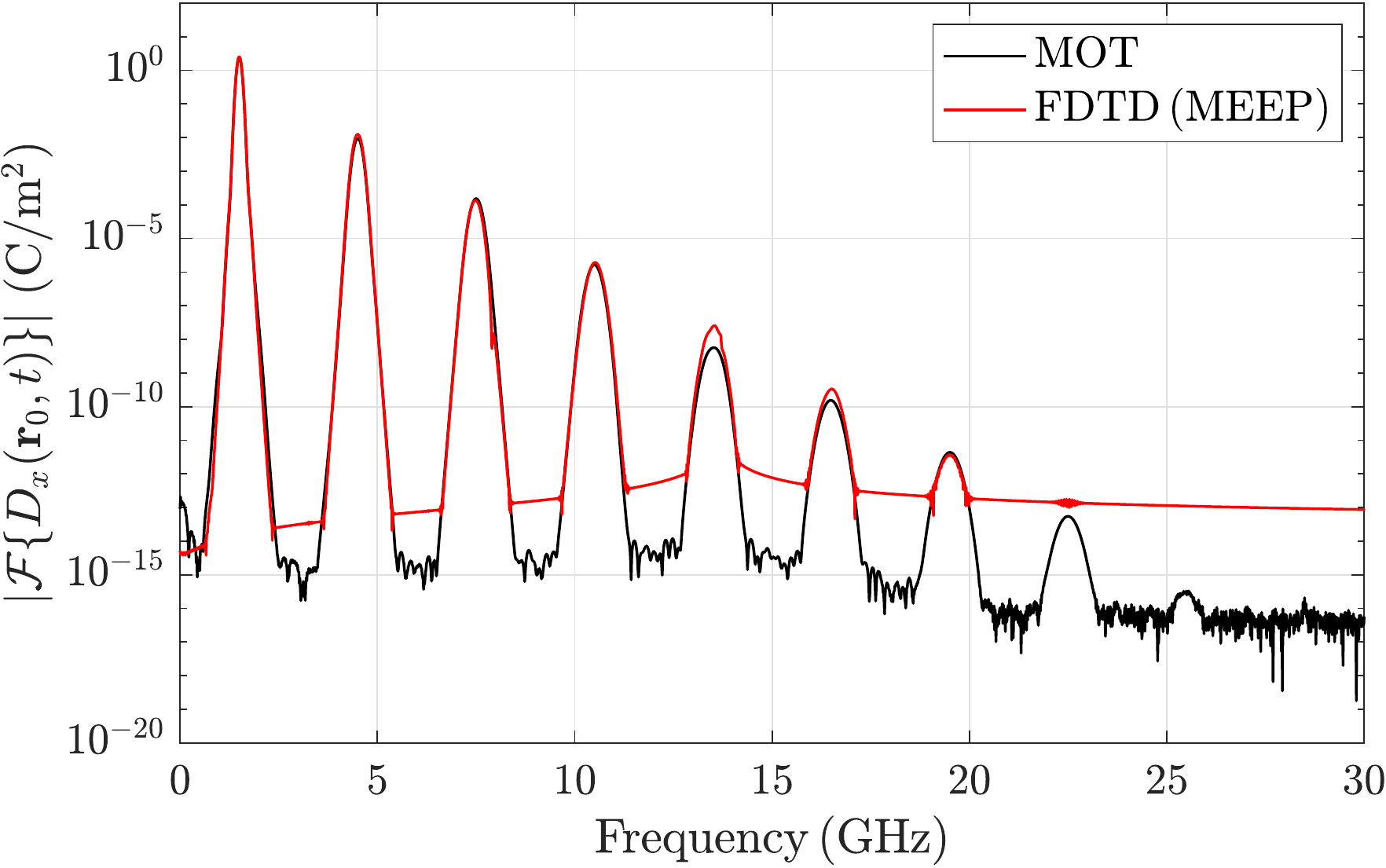}}\hspace{0.5cm}
\subfigure[]{\includegraphics[width=0.46\columnwidth]{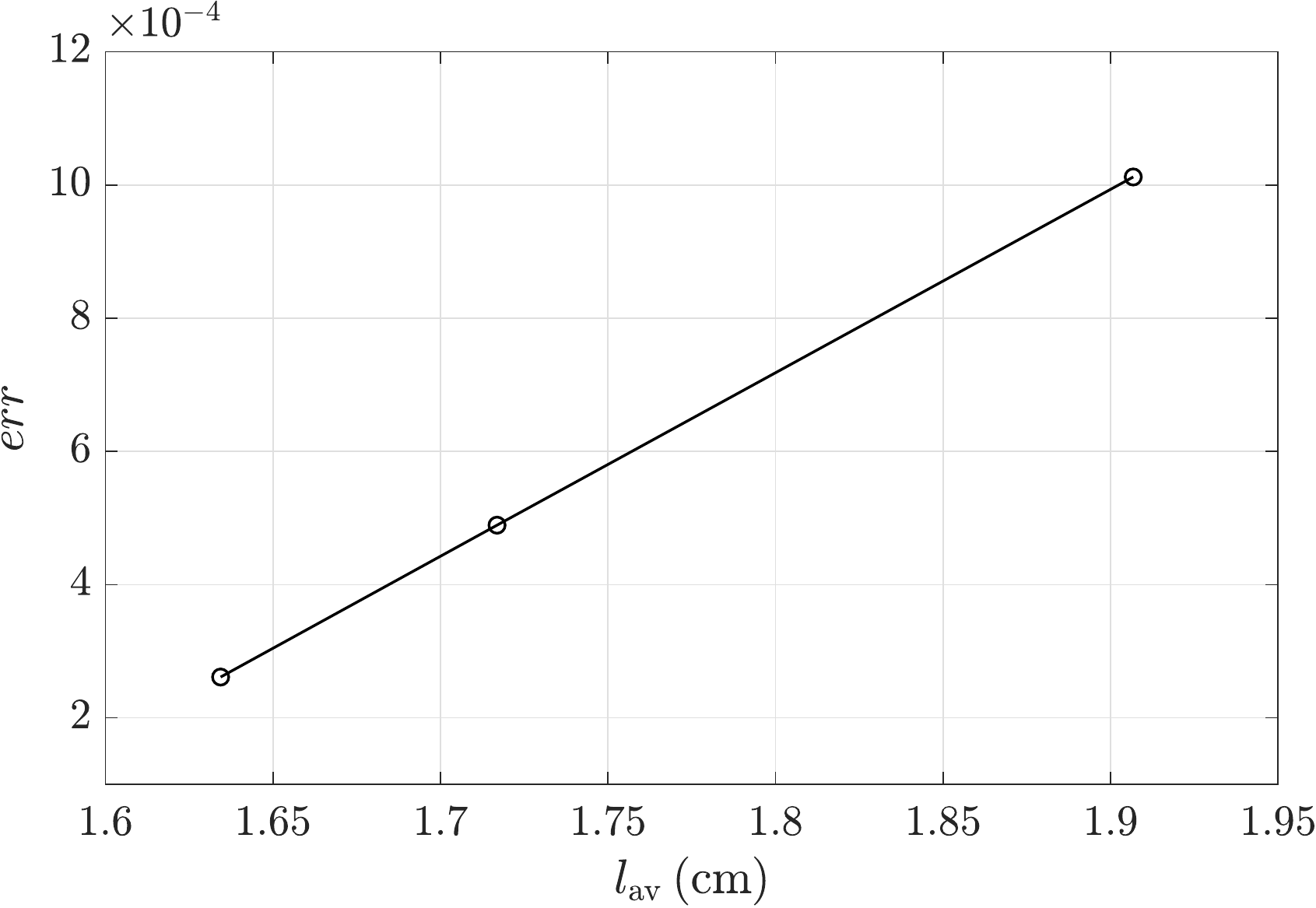}}
\caption{Scattering from a nonlinear cube. (a) $x$-component of $\mathbf{D}(\mathbf{r}_0,t)$ computed by the proposed solver (MOT) and the FDTD-based solver MEEP at the center of the cube [$\mathbf{r}_0=(0,0,0)$]. (b) Zoomed version of (a) in the time range $[32.5, 41.5]\,\mathrm{ns}$. (c) Fourier transform of the $x$-component of $\mathbf{D}(\mathbf{r}_0,t)$ computed by the proposed solver (MOT) and the FDTD-based solver MEEP. (d) Convergence in $err$ defined by~\eqref{eq:errorcalc} with increasing mesh density (decreasing average edge length $l_{\mathrm{av}}$).}
\label{fig:nonlinear_cube}
\end{figure*}
\begin{figure*}[t!]
\centering
\subfigure[]{\includegraphics[width=0.47\columnwidth]{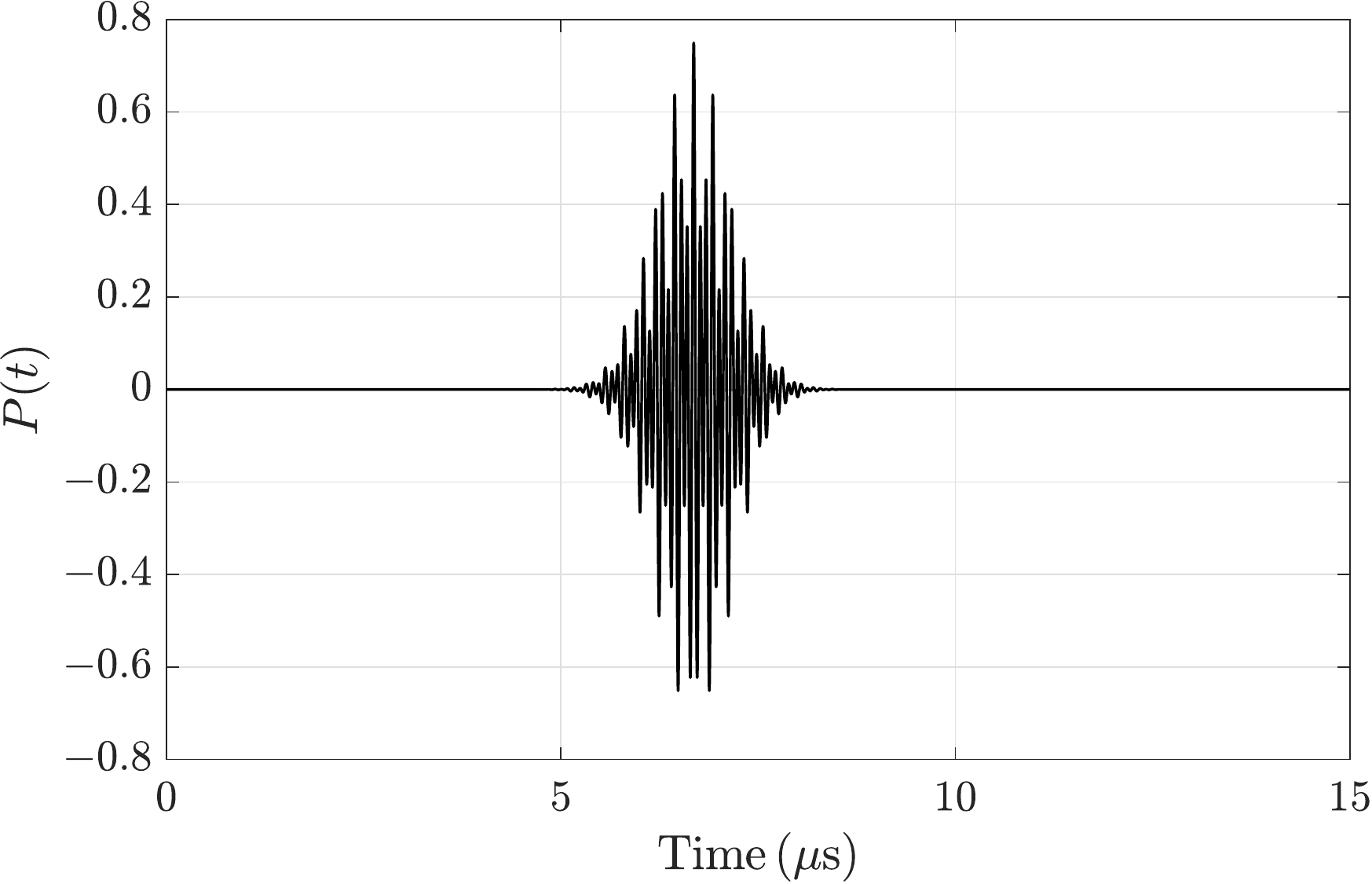}}\hspace{0.25cm}
\subfigure[]{\includegraphics[width=0.48\columnwidth]{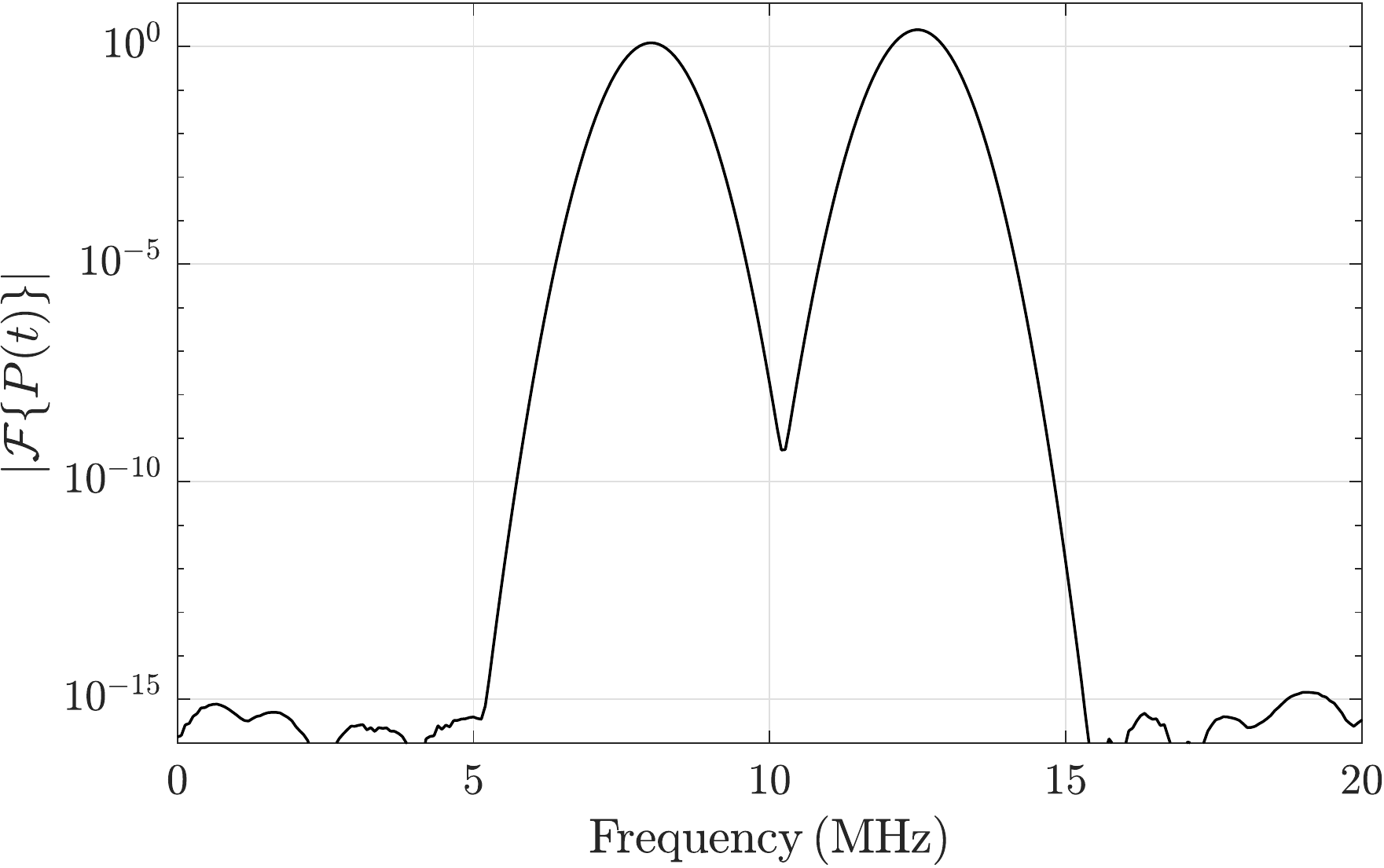}}
\subfigure[]{\includegraphics[width=0.47\columnwidth]{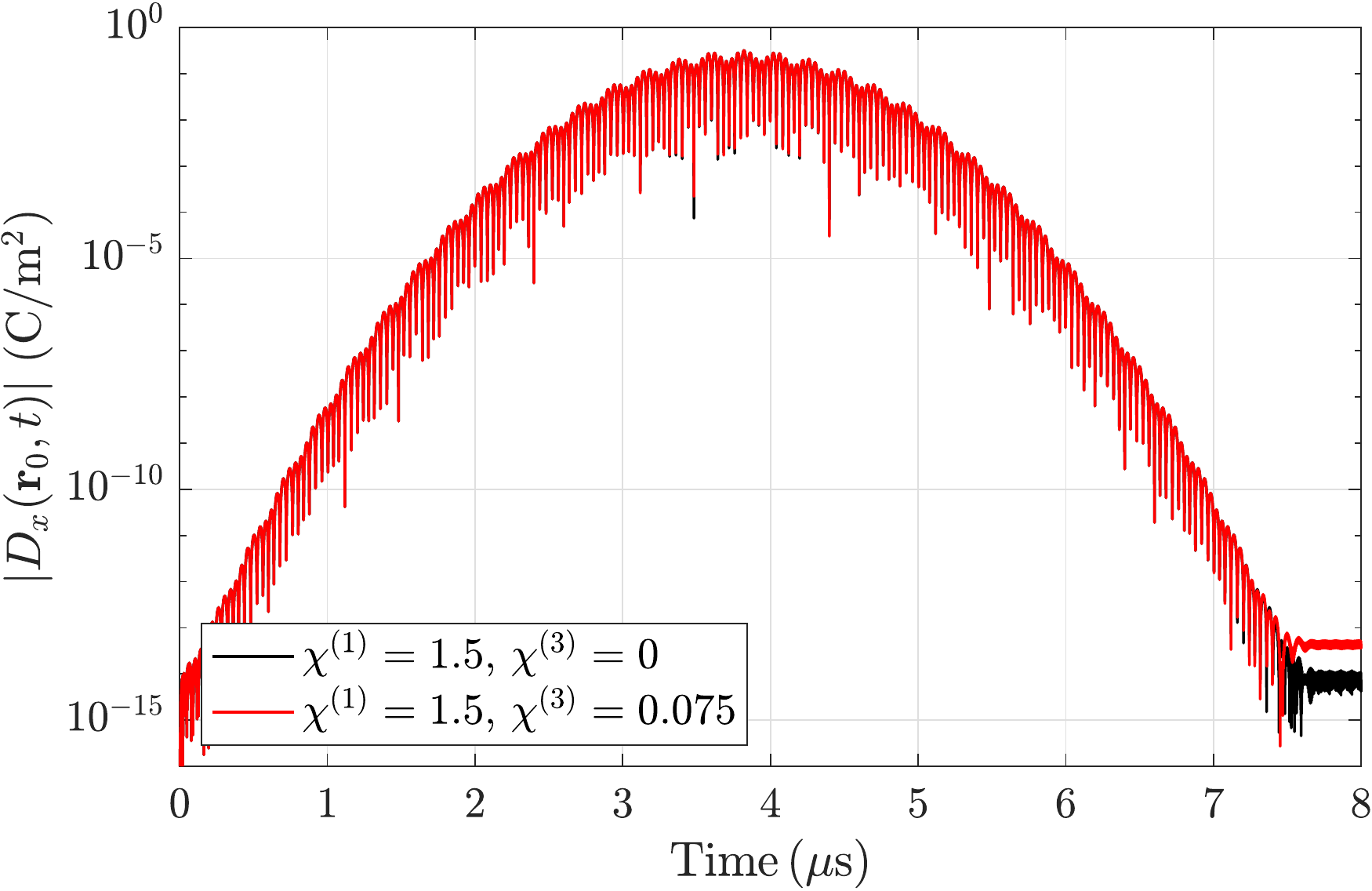}}\hspace{0.25cm}
\subfigure[]{\includegraphics[width=0.48\columnwidth]{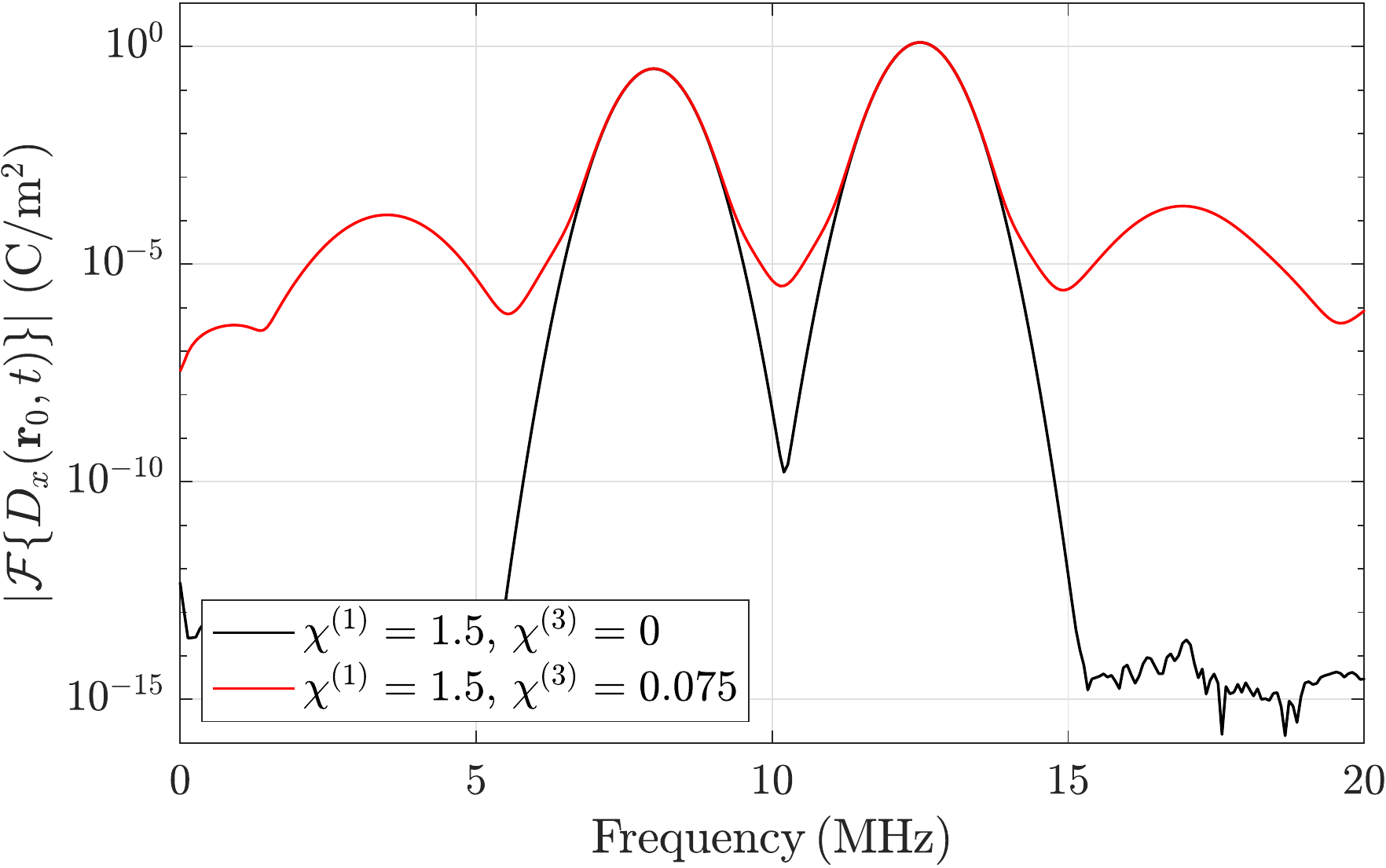}}
\caption{Four-wave mixing frequency conversion. (a) The time dependence of the plane wave excitation, $P(t)$ given by~\eqref{eq:4wave-mixing}, and (b) its Fourier transform. (c) $x$-component of $\mathbf{D}(\mathbf{r}_0,t)$ computed by the proposed solver at the center of the sphere in the first (linear sphere with $\chi^{(1)}=1.5$, $\chi^{(3)}=0$) and the second (nonlinear sphere with $\chi^{(1)}=1.5$, $\chi^{(3)}=0.075$) simulations [$\mathbf{r}_0=(0,0,0)$]. (d) Fourier transform of the $x$-component of $\mathbf{D}(\mathbf{r}_0,t)$ computed in the two simulations.}
\label{fig:4wave-mixing}
\end{figure*}

In the second example, electromagnetic scattering from a nonlinear cube is analyzed using the proposed solver. The cube is centered at the origin and has an edge of length $0.1\,\mathrm{m}$. The coefficients of the permittivity function of the cube are $\chi^{(1)}=2$ and $\chi^{(3)}=0.01$. The time dependence of the plane wave excitation is given by~\eqref{eq:linear_sphere} with $f_{0}=1498.96\,\mathrm{MHz}$, $f_{\mathrm{bw}}=149.90\,\mathrm{MHz}$, $\sigma=3.1853\,\mathrm{ns}$, and $t_{\mathrm{p}}=10 \sigma$.

$\mathbf{E}(\mathbf{r}, t)$ and $\mathbf{D}(\mathbf{r}, t)$ induced inside the cube are discretized using $N^{\mathrm{E}}=7\,596$ and $N^{\mathrm{D}}=4\,002$ spatial basis functions, respectively. The MOT scheme is executed for $N_{\mathrm{t}}=6\,000$ time steps with $\Delta t=13.343\,\mathrm{ps}$. The SOR coefficient in~\eqref{eq:step7_2} is selected as $\alpha=0.3$.

To verify the results of the proposed solver, this scattering scenario is also analyzed using the open-source FDTD-based solver MEEP~\cite{oskooi2010Meep}. The dimension of the FDTD computation domain is $1\,\mathrm{m} \times 1\,\mathrm{m} \times 2\,\mathrm{m}$ and the thickness of the perfectly matched layer (PML) is $0.3\,\mathrm{m}$. The computation domain and the PML are discretized using Yee cells of dimension $5\,\mathrm{mm}$ and the time discretization uses a time step of size $\Delta t=8.33\,\mathrm{ps}$.

Fig.~\ref{fig:nonlinear_cube}(a) compares the $x$-component of $\mathbf{D}(\mathbf{r}, t)$ computed by the proposed solver and MEEP at the center of the cube. Fig.~\ref{fig:nonlinear_cube}(b) zooms to the time range $[32.5,41.5]\,\mathrm{ns}$ of the curves in Fig.~\ref{fig:nonlinear_cube}(a). Both figures show that the results agree very well for $t>12\,\mathrm{ns}$ and $t<50\,\mathrm{ns}$. The discrepancy between the results outside this range is because MEEP cannot capture the solution accurately when it is small. Ideally, both solutions should decay to zero. From this perspective, one can argue that the accuracy of the proposed solver is actually higher than MEEP since the late-time solution obtained by the proposed solver reaches a lower level~\cite{chen2019,sayed2020,chen2021,chen2022}. This is further investigated by studying the Fourier transform of the solutions as described next.  

Fig.~\ref{fig:nonlinear_cube}(c) compares the Fourier transform of the solutions in Fig.~\ref{fig:nonlinear_cube}(a) in the frequency range $f \in[0, 30.0]\,\mathrm{GHz}$. The figure clearly shows that several higher-order harmonics are generated due to the nonlinearity of the dielectric permittivity. Fourier transformed solutions match well up to frequencies where the numerical error in the MEEP solution becomes high enough to significantly effect the solution. Indeed, the mismatch between the two Fourier transformed solutions around the fifth harmonic is on the order of the difference in the levels of the two solutions in the late time as shown in Fig.~\ref{fig:nonlinear_cube}(a).

Next, it is demonstrated that the solution obtained by the proposed solver converges with increasing mesh density. Four different meshes with average edge length $l_{\mathrm{av}}=\{1.896$, $1.716$, $1.634$, $1.553\}\,\mathrm{cm}$ are considered. This results in $N^{\mathrm{E}}=\{5\,860$, $7\,596$, $9\,108$, $10\,784\}$ and $N^{\mathrm{D}}=\{3\,110$, $4\,002$, $4\,812$, $5\,668 \}$ for four simulations, respectively. All other parameters are kept the same. After each time domain simulation, Fourier transformed solutions are used to compute the scattered electric far-field $\mathbf{E}^{\mathrm{sca}}_{{2n-1},k}$ for $\phi=0$ and $\theta = k\Delta\theta$, $\Delta\theta=1.0^{\circ}$, $k=1,2,\ldots,180$ at the center frequencies of the first ten harmonics $(2n-1)f_{0}$, $n=1,2\ldots,10$. The following error is used as a measure of the convergence:
\begin{equation}
\label{eq:errorcalc}
err=\sqrt{\frac{1}{10} \sum_{n=1}^{10} \frac{1}{180} \sum_{k=1}^{180}\Big|\mathbf{E}_{2 n-1,k}^{\mathrm {sca }}-\mathbf{E}_{2 n-1,k}^{\mathrm {sca,ref }}\Big|^2}
\end{equation}
where $\mathbf{E}^{\mathrm{sca,ref}}_{{2n-1},k}$ is the scattered electric far-field computed by the simulation with the densest mesh. Fig.~\ref{fig:nonlinear_cube}(d) plots $err$ versus $l_{\mathrm{av}}$ and shows that the solution obtained by the proposed solver converges with increasing mesh density.

\begin{figure*}[t!]
\centering
\subfigure[]{\includegraphics[width=0.46\columnwidth]{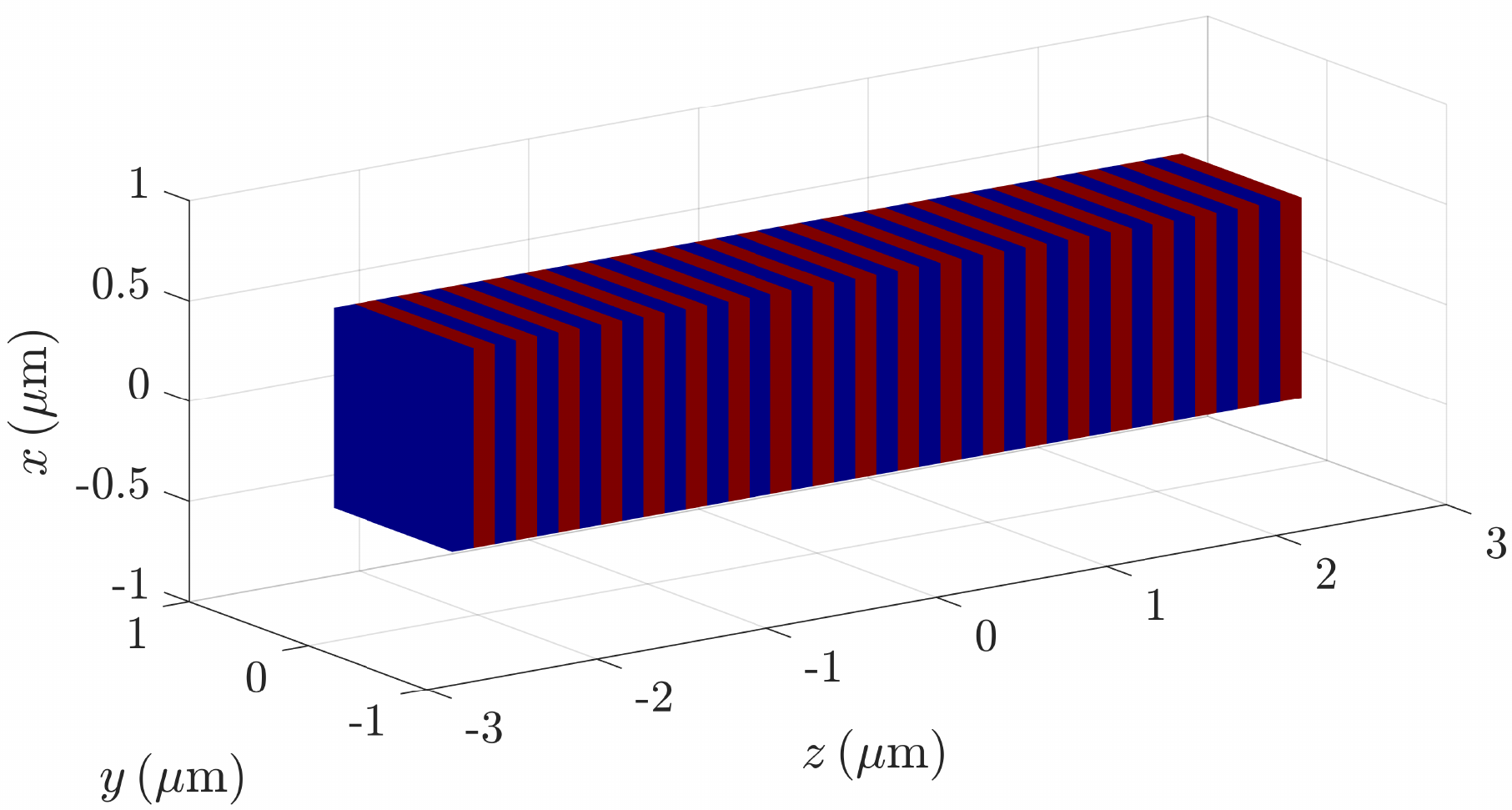}}\hspace{0.25cm}
\subfigure[]{\includegraphics[width=0.48\columnwidth]{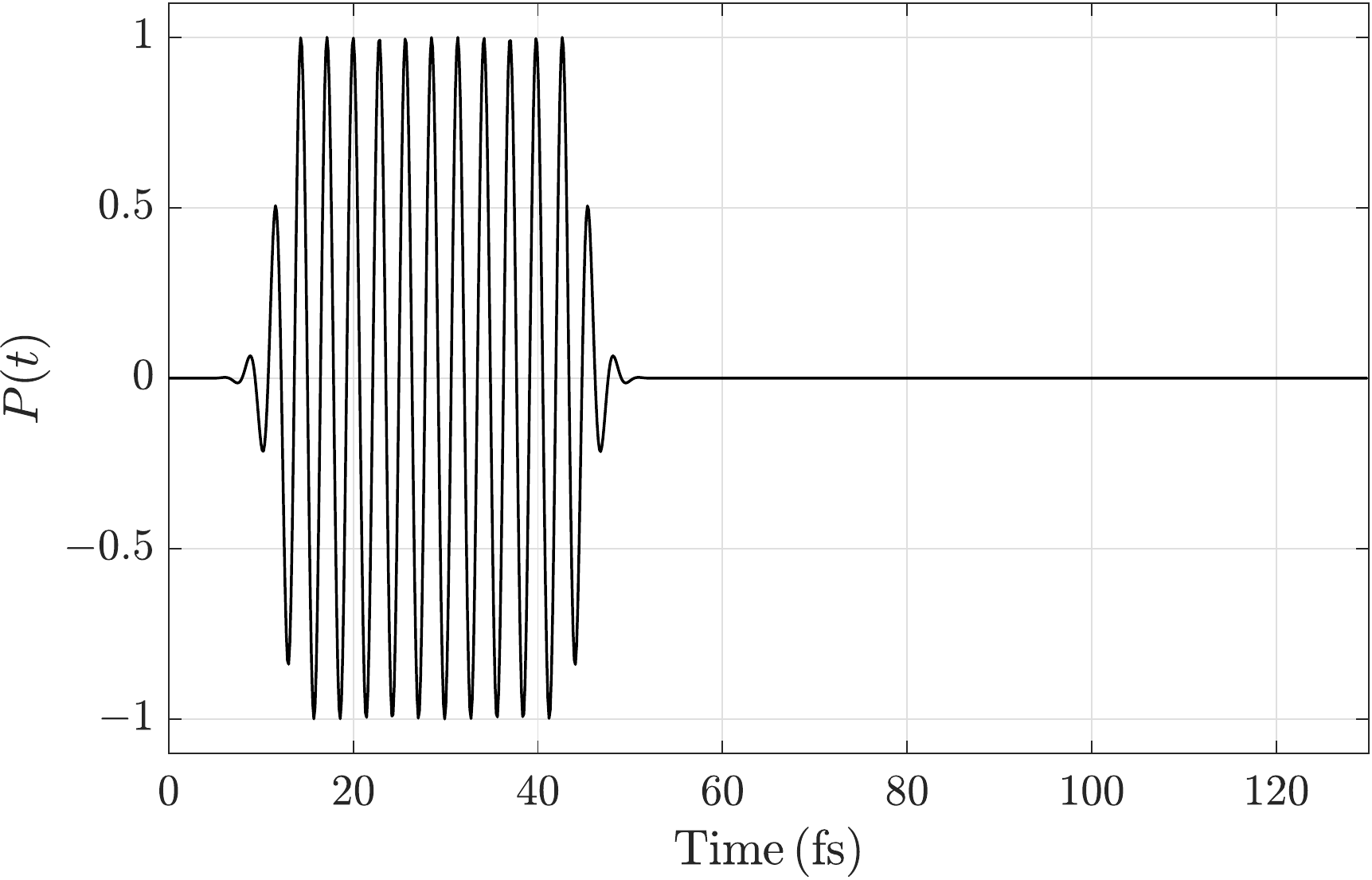}}
\subfigure[]{\includegraphics[width=0.46\columnwidth]{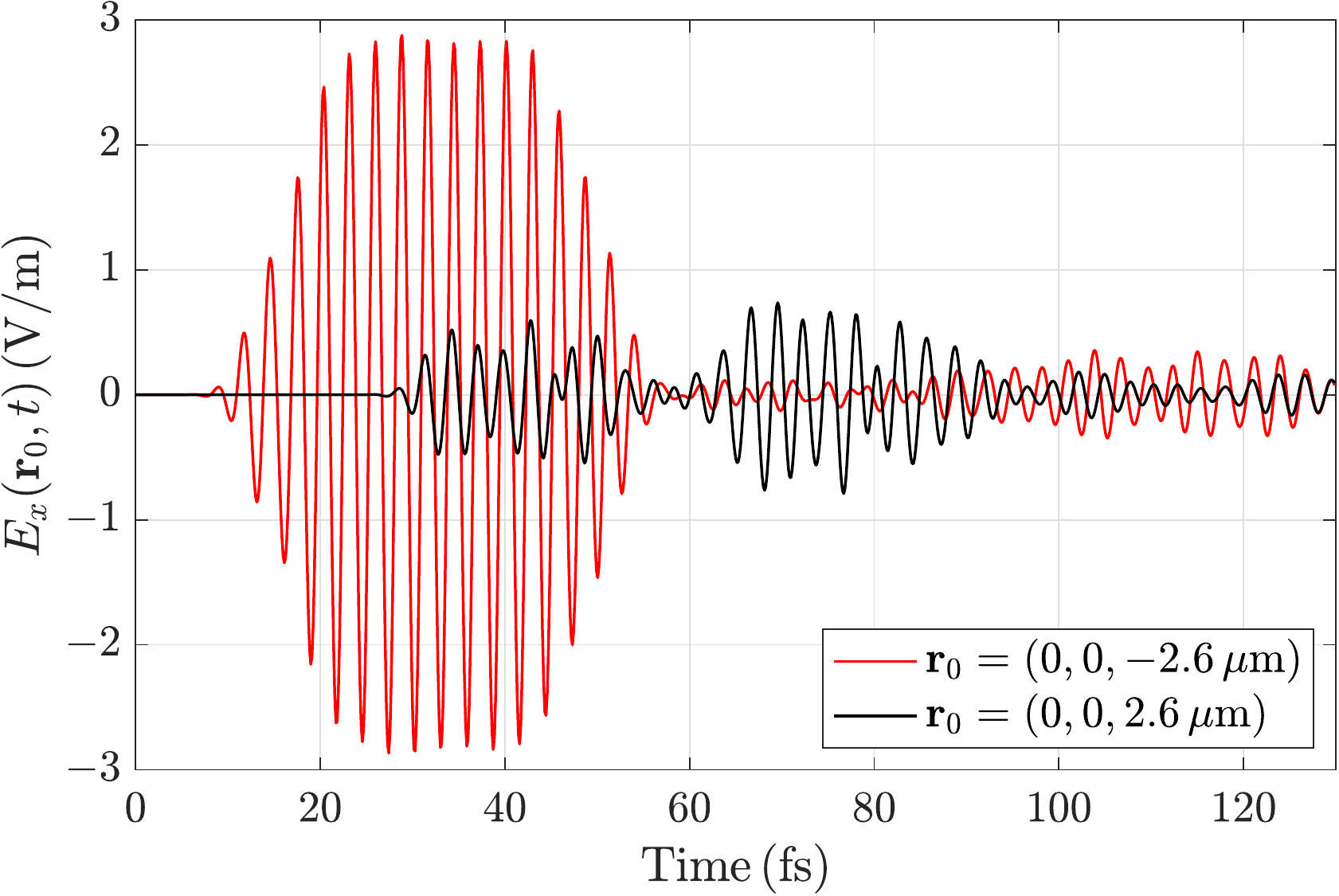}}\hspace{0.25cm}
\subfigure[]{\includegraphics[width=0.48\columnwidth]{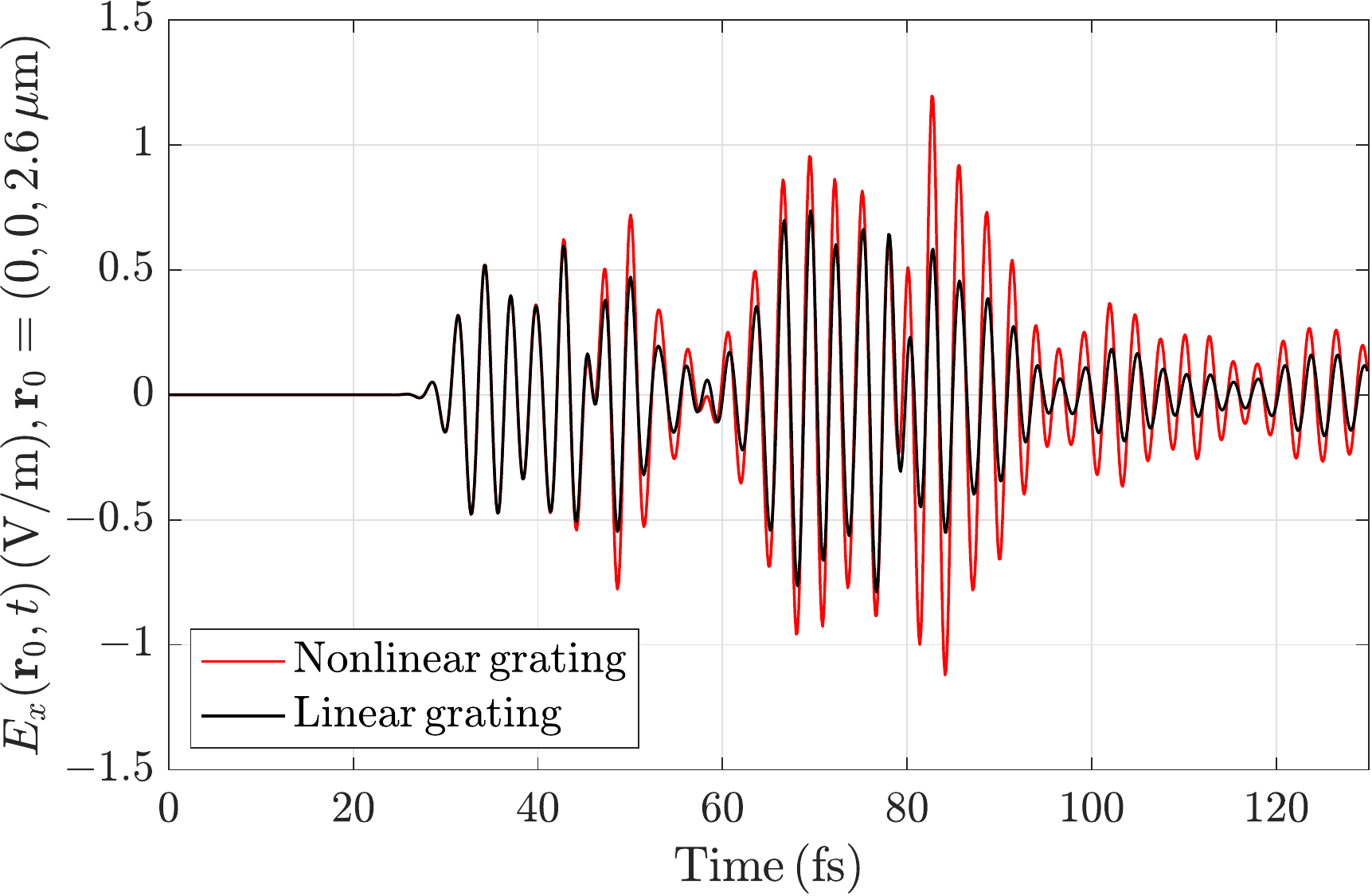}}
\caption{Transmission through a Bragg grating. (a) Description of the grating geometry. (b) The time dependence of the plane wave excitation, $P(t)$ given by~\eqref{eq:nonlinear_bragg}. (c) $x$-component of $\mathbf{E}(\mathbf{r}, t)$ computed at the feeding end [$\mathbf{r}_0=(0,0,-2.6\,\mu \mathrm{m})$] and the trailing end [$\mathbf{r}_0=(0,0,2.6\,\mu \mathrm{m})$] in the first simulation, where both layers are linear. (d) $x$-component of $\mathbf{E}(\mathbf{r}, t)$ at the trailing end [$\mathbf{r}_0=(0,0,2.6\,\mu \mathrm{m})$] computed by the proposed solver in the first (both layers are linear) and the second (one layer is linear, the other one is nonlinear) simulations.}
\label{fig:nonlinear_bragg}
\end{figure*}
\subsection{Four-wave Mixing}\label{sec:4wave-mixing}
In this example, four-wave mixing frequency conversion of electromagnetic fields~\cite{fujii2005} is analyzed using the proposed method. The scatterer is a sphere that is centered at the origin and has a radius of length $1.0\,\mathrm{m}$. Two simulations are carried out to clearly demonstrate that the nonlinearity results in four-wave mixing. In the first simulation, the sphere is linear and the coefficients of its permittivity function are $\chi^{(1)}=1.5$ and $\chi^{(3)}=0$. In the second simulation, the sphere is nonlinear and the coefficients of its permittivity function are $\chi^{(1)}=1.5$ and $\chi^{(3)}=0.075$. In both simulations, the time dependence of the plane wave excitation in~\eqref{eq:excitation} is a sum of two modulated Gaussian pulses and is expressed as
\begin{equation}
\label{eq:4wave-mixing}
\begin{aligned}
P(t)=&\left(0.25\cos \left[2 \pi f_{1}\left(t-t_{\mathrm{p}}\right)\right]\right.\\
&\left.+0.5\cos \left[2 \pi f_{2}\left(t-t_{\mathrm{p}}\right)\right]\right)e^{-(t-t_{\mathrm{p}})^2/(2\sigma^2)}
\end{aligned}
\end{equation}
where $f_{1}=8.0\,\mathrm{MHz}$, $f_{2}=12.5\,\mathrm{MHz}$, $f_{\mathrm{bw}}=1.0\,\mathrm{MHz}$, $\sigma = 0.4775\,\mu\mathrm{s}$, and $t_{\mathrm{p}}=14 \sigma$. $P(t)$ and its Fourier transform are plotted in Figs.~\ref{fig:4wave-mixing}(a) and (b), respectively. Both simulations use the same mesh, and $\mathbf{E}(\mathbf{r}, t)$ and $\mathbf{D}(\mathbf{r}, t)$ induced inside the sphere are discretized using $N^{\mathrm{E}}=4\,444$ and $N^{\mathrm{D}}=2\,686$ spatial basis functions, respectively. The simulations are executed for $N_{\mathrm{t}}=15\,000$ with $\Delta t=1.0\,\mathrm{ns}$. The SOR coefficient in~\eqref{eq:step7_2} is selected as $\alpha=0.3$.

Fig.~\ref{fig:4wave-mixing}(c) compares the $x$-component of $\mathbf{D}(\mathbf{r}, t)$ computed by the proposed solver at the center of the sphere in the first ($\chi^{(1)}=1.5$, $\chi^{(3)}=0$) and the second ($\chi^{(1)}=1.5$, $\chi^{(3)}=0.075$) simulations. The figure shows that both solutions are stable. Fig.~\ref{fig:4wave-mixing}(d) compares the Fourier transform of these solutions in the frequency range $f \in[0,20]\,\mathrm{MHz}$. The two peaks observed at frequencies $f=f_{1}$ and $f=f_{2}$ in both solutions match the peaks in the Fourier transform of the excitation pulse shown in Fig.~\ref{fig:4wave-mixing}(b) ($f_1$ and $f_2$ are the modulation frequencies of the two Gaussian pulses added in~\eqref{eq:4wave-mixing}). However, the Fourier transform of the solution in the second simulation has two extra peaks at frequencies $f=2 f_{1}-f_{2}$ and $f=2 f_{2}-f_{1}$. These peaks are observed in the electromagnetic response because of the four-wave mixing frequency conversion generated as a result of the $\chi^{(3)}$-term in the permittivity function (i.e., Kerr nonlinearity)~\cite{fujii2005}.

\subsection{Nonlinear Bragg Grating}\label{sec:nonlinear_bragg}
In the last example, electromagnetic scattering from a nonlinear Bragg grating is analyzed using the proposed solver. The grating consists of 40 alternating layers of dielectric materials with permittivity functions $\varepsilon_{1}(\mathbf{r},t,\mathbf{E})=\varepsilon_{0}[\chi_1^{(1)}+\chi_1^{(3)}|\mathbf{E}(\mathbf{r}, t)|^{2}]$ and $\varepsilon_{2}(\mathbf{r},t,\mathbf{E})=\varepsilon_{0}[\chi_2^{(1)}+\chi_2^{(3)}|\mathbf{E}(\mathbf{r}, t)|^{2}]$ [Fig.~\ref{fig:nonlinear_bragg}(a)]. All layers have the same thickness and the dimension of the whole grating is $1\,\mu\mathrm{m} \times 1\,\mu\mathrm{m} \times 5\,\mu\mathrm{m}$. Two simulations are carried out. In the first simulation, both layers are linear with 
$\{\chi_1^{(1)}=2.25, \chi_1^{(3)}=0\}$ and $\{\chi_2^{(1)}=4.5, \chi_2^{(3)}=0\}$. In the second simulation, first layer is linear with $\{\chi_1^{(1)}=2.25, \chi_1^{(3)}=0\}$ but the second layer is nonlinear with $\{\chi_2^{(1)}=4.5, \chi_2^{(3)}=-0.06\}$. In both simulation, the time dependence of the plane wave excitation in~\eqref{eq:excitation} is expressed as
\begin{equation}
\label{eq:nonlinear_bragg}
P(t)=\left\{\begin{aligned}
&\cos \left[2 \pi f_{0}\left(t-t_{1}\right)\right] e^{-\left(t-t_{1}\right)^{2} / 2 \sigma^{2}}, t<t_{1} \\
&\cos \left[2 \pi f_{0}\left(t-t_{1}\right)\right], t_{1} \leq t<t_{2} \\
&\cos \left[2 \pi f_{0}\left(t-t_{2}\right)\right] e^{-\left(t-t_{2}\right)^{2} / 2 \sigma^{2}}, t \geq t_{2}
\end{aligned}\right.
\end{equation}
where $f_{0}=353.0\,\mathrm{THz}$, $f_{\mathrm{bw}}=200.0\,\mathrm{THz}$, $\sigma=2.3873\,\mathrm{fs}$, $t_{1}=6 \sigma$, and $t_{2}=17.87 \sigma$. Fig.~\ref{fig:nonlinear_bragg}(b) plots $P(t)$. Both simulations use the same mesh, and $\mathbf{E}(\mathbf{r}, t)$ and $\mathbf{D}(\mathbf{r}, t)$ induced inside the grating are discretized using $N^{\mathrm{E}}=22\,680$ and $N^{\mathrm{D}}=13\,787$ spatial basis functions, respectively. The simulations are executed for $N_{\mathrm{t}}=1\,000$ with $\Delta t=0.13\,\mathrm{fs}$. The SOR coefficient in~\eqref{eq:step7_2} is selected as $\alpha=0.4$.

Fig.~\ref{fig:nonlinear_bragg}(c) compares the $x$-component of $\mathbf{E}(\mathbf{r}, t)$ computed at the feeding end [$\mathbf{r}_0=(0$, $0$, $-2.6\,\mu \mathrm{m})$] and the trailing end [$\mathbf{r}_0=(0,0,2.6\,\mu \mathrm{m})$] in the first simulation, where both layers are linear. The figure clearly shows that the electric field at the trailing end is much smaller than the one at the feeding end. This is due to the fact that the linear Bragg grating has a stop band between $300\,\mathrm{THz}$ and $370\,\mathrm{THz}$~\cite{sarris2011} and the significant part of excitation's power is within this stop band. Fig.~\ref{fig:nonlinear_bragg}(d) compares the $x$-component of $\mathbf{E}(\mathbf{r}, t)$ at the trailing end [$\mathbf{r}_0=(0,0,2.6\,\mu \mathrm{m})$] computed by the proposed solver in the first (both layers are linear) and the second (one layer is linear, the other one is nonlinear) simulations. The figure shows that with the introduction of the nonlinearity, the electric field at the trailing end is enhanced. This can be explained by the fact that the stop band for the linear Bragg grating can be partially closed by introducing a negative Kerr nonlinearity~\cite{sarris2011}. 


\section{Conclusion}\label{sec:conc}
An explicit MOT-based TD-EFVIE solver is developed to analyze electromagnetic scattering from dielectric objects with Kerr nonlinearity. The nonlinear constitutive relation that relates electric flux and electric field induced in the scatterer is used as an auxiliary equation that complements TD-EFVIE. Discretizing TD-EFVIE using SWG functions yields a system of ODEs in time-dependent SWG expansion coefficients. This system is integrated in time using a $PE(CE)^m$ scheme to obtain the expansion coefficients of the electric field. Similarly, the nonlinear constitutive relation and its inverse obtained using the Padé approximant are discretized using SWG functions. The resulting matrices are used to carry out explicit updates of electric field and electric flux expansion coefficients at the predictor (PE) and the corrector (CE) stages. This approach produces an explicit MOT scheme that does not call for any Newton-like nonlinear solver but only requires solution of sparse and well-conditioned Gram matrix systems at every step. These solutions are done very efficiently using an iterative solver. 

The accuracy and the applicability of the explicit MOT-based TD-EFVIE solver are demonstrated using several numerical examples. These results clearly show that the proposed method is more accurate than FDTD that is traditionally used for analyzing electromagnetic scattering from nonlinear objects.

Extension of the proposed scheme to analyze nonlinear problems involving high-contrast dielectric scatterers is underway.

\appendices








\end{document}